\documentclass[a4paper,11pt]{article}
\pdfoutput=1 
\usepackage{jcappub} 

\usepackage[T1]{fontenc} 
\usepackage{aas_macros}
\usepackage{natbib}
\usepackage{graphicx}	
\usepackage{amsmath}	
\usepackage{amssymb}	
\usepackage{mathptmx}
\usepackage{txfonts} 
\usepackage{subcaption} 
\usepackage{pdflscape}
\usepackage{rotating}
\usepackage{multirow} 
\usepackage{soul,xcolor}
\usepackage{color}
\definecolor{darkgreen}{rgb}{0,0.5,0}
\definecolor{mag}{rgb}{0.79,0.08,0.48}
\definecolor{darkblue}{rgb}{0,0.2,0.8}
\definecolor{cyan}{rgb}{0,0.8,0.8}

\begin{document}
\setstcolor{mag}
\title{Galaxy-lens determination of $H_0$: constraining density slope in the context of the mass sheet degeneracy}


\author[a,1]{M. Gomer,\note{Corresponding author.}}
\author[a]{L. L. R. Williams,}


\affiliation[a]{University of Minnesota\\ 116 Church Street SE, Minneapolis MN, 55455, USA}

\emailAdd{gomer011@umn.edu}
\emailAdd{llrw@umn.edu}

\abstract{Gravitational lensing offers a competitive method to measure $H_0$ with the goal of 1\% precision. A major obstacle comes in the form of lensing degeneracies, such as the mass sheet degeneracy (MSD), which make it possible for a family of density profiles to reproduce the same lensing observables but return different values of $H_0$. The modeling process artificially selects one choice from this family, potentially biasing $H_0$. The effect is more pronounced when the profile of a given lens is not perfectly described by the lens model, which will always be the case to some extent. To explore this, we quantify the bias and spread in $H_0$ by creating quads from two-component mass models and fitting them with a power-law ellipse+shear model. We find that the bias does not correspond to the estimate one would calculate by transforming the profile into a power law near the image radius. We also emulate the effect of including stellar kinematics by performing fits where the slope is  constrained to the true value. Informing the fit using the true value near the image radius can introduce substantial bias (0-23\% depending on the model). We confirm using Jeans arguments that kinematic constraints can result in a biased value of $H_0$, though the degree of bias depends on the region kinematic modeling probes in specific lenses. We conclude that lensing degeneracies manifest through commonplace modeling approaches in a more complicated way than is assumed in the literature. If stellar kinematics incorrectly break the MSD, their inclusion may introduce more bias than their omission.
}

\maketitle
\flushbottom


\section{Introduction}
       
Robust determination of the Hubble constant is one of the most sought-after goals in cosmology. Over the years, 
increasingly precise measurements of temperature anisotropies in the 
cosmic microwave background (CMB) have recovered values of $H_0$ with smaller and smaller uncertainties. 
At present, the most precise CMB constraints come from the Planck mission, which found $H_0=67.36 \pm 0.54$ km s$^{-1}$Mpc$^{-1}$ (0.7\% uncertainty),
assuming $\Lambda$CDM cosmology \citep{Planck18}. 
Baryon Acoustic Oscillation (BAO) results from the Dark Energy Survey are broadly consistent with CMB results \citep{Abbott17}.

Meanwhile, standard candle observations using Type Ia supernovae and Cepheid variables provide a direct distance measurement to 
faraway galaxies, allowing $H_0$ to be measured directly rather than recovered from a model with many parameters \citep{Riess16}. 
The tradeoff is that this method is dependent on the calibration of these standard candles, 
where any uncertainties in local measurements propogate to farther measurements.
This method has been able to compete with the precision of CMB observations and, through improvements in the calibration, has presently 
determined value of $H_0=74.03 \pm 1.42$ km s$^{-1}$Mpc$^{-1}$ (1.91\% uncertainty) \citep{Riess19}. 

Tension exists between these two methods at the $4.4\sigma$ level. The cause of this tension is unknown at present. These two methods
compare the directly-measured local value of $H_0$ to the most distant possible determination at the time of recombination, meaning they
probe the expansion of the universe from one end to another. It might turn out that the prevailing 
model is more complicated than $\Lambda$CDM, hinting at new physics beyond the standard model or general relativity, 
perhaps through time-dependent dark energy or some other mechanism \citep{Riess16}. 
On the other hand, it might turn out that the uncertainties of these two methods are missing some source of 
systematic error, and are thus underestimated. 
If the Milky Way resides within a local void, the measured value of $H_0$ will be systematically biased with respect to the true value, although
at present it does not seem that this effect would be sufficient to resolve the tension \citep{Fleury17,DArcy19}.      
Perhaps the answer lies in the standard candle calibration-- \citet{Freedman19} recently found that replacing the Cepheid variable calibration with the
Tip of the Red Giant Branch (TRGB) distance indicator results in a lower value of 
$H_0=69.8 \pm 0.8$ $(\pm1.1\%$ stat$) \pm 1.7$ $(\pm2.4\%$ sys$)$ km s$^{-1}$Mpc$^{-1}$, only $1.2\sigma$ away from the CMB result.       
It might turn out to be random chance that the methods disagree and as more data is collected they may converge to the same value.
To diagnose the existence of the tension between these methods, an additional independent method would be exceedingly useful. 

Strong gravitational lensing offers this independent method. If a variable source is multiply imaged, 
the difference in arrival time between the images offers a measurement of the time delay distance,
$D_{\Delta t}=(1+z_d)\frac{D_d D_s}{D_{ds}} \propto \frac{1}{H_0}$ \citep{Refsdal64}.
If one has an accurate model of the lensing potential for the mass distribution of the lens, it is straightforward to measure $H_0$ from
such information \citep{Schechter97,Suyu10,HC1}. 
The challenge is to precisely determine the time delays and lensing potential. 

At the cluster scale, multiple sources provide additional constraints to the potential and allow one 
to mitigate the effects of the mass-sheet degeneracy (see Section \ref{ssec:degeneracies}). Using parametric models which implicitly assume
that the galaxies within the cluster have similar mass profiles to isolated galaxies in equilibrium, constraints on $H_0$ have been estimated at the
6\% level, with 40\% uncertainty on $\Omega_m$ \citep{Grillo20}.
When this assumption is relaxed through the use of free-form modeling, the lensing potential is considerably
more complicated, producing larger uncertainties in $H_0$ \citep{Williams19}. 
Because of this, the strongest constraints will likely come from the scale of individual galaxies, which will be the focus of this paper.

Improvements in the method over time have enabled constraints on $H_0$ to be placed at the 7\% level using a single system \citep{Suyu10,Suyu14}. 
Further improvements can be gained by combining constraints using multiple systems to average over variations between individual lenses.
The tightest constraints from this method come from the state-of-the-art H0LiCOW ($H_0$ Lenses in COSMOGRAIL's Wellspring) program \citep{HC1}.
H0LiCOW gets time delays from the COSMOGRAIL (COSmological MOnitoring of GRAvItational Lenses) program \citep{Courbin04,Bonvin16},
which has been monitoring light curves of multiply imaged systems for 15 years to date, measuring 
time delays within 1-3\% uncertainty \citep{RKumar13,Tewes13}. 
H0LiCOW models lenses using image positions, fluxes, and time delays, as well as stellar kinematics of the lens galaxy \citep{HC4}.
The analysis incorporates a variety of effects, such as inclusion of nearby group members \citep{HC2} and an estimation of
line-of-sight external convergence \citep{HC3}.
Most recently, a blind combined analysis of six systems yielded a measurement of 
$H_0=73.3 \substack{+1.7 \\ -1.8}$ km s$^{-1}$Mpc$^{-1}$ (2.4\% uncertainty) \citep{HC13}.

In order to provide insight into the 
nature of the $H_0$ tension, the uncertainty in the method must be competitive with the existing methods. The ambitious goal of the 
community is to reduce the uncertainties of the time delay method to 1\% \citep{HC1}. \citet{HC5} outline four actions which must be taken
in order to reach such high precision: 1. Enlarge the sample, 2. Improve the lens model accuracy, 3. Improve the mass calibration through
spatially resolved kinematics, and 4. Increase the efficiency of time delay measurement techniques. While the other actions are certainly
important, the focus of this paper will be on the second and third: improving the lens model accuracy and studying the role of kinematic information. 
As \citet{HC5} put it, ``as
the number of systems being analysed grows, random uncertainties in the cosmological parameters will fall, and residual systematic
uncertainties related to degeneracies inherent to gravitational lensing will need to be investigated in more detail."  Put another way,
the statistical scatter due to a small sample size will decrease with time, but any biases inherent to lens modeling will not go away,
potentially offsetting the recovered value from the true value. It is crucial that all biases intrinsic to the modeling process are 
carefully accounted for.

\subsection{Effect of lensing degeneracies}\label{ssec:degeneracies}
The source of the problem comes from lensing degeneracies, where the same observables can be recovered with multiple different 
lens models. There exist many types of degeneraries, the most famous of which is the exact mass sheet degeneracy (MSD) \citep{Falco85}, where image
positions and relative fluxes are left unchanged by a rescaling of the profile normalization and the corresponding 
introduction of a uniform convergence.
\begin{equation}
\kappa_{\lambda}(\vec{x})=\lambda\kappa(\vec{x})+(1-\lambda)
\end{equation}

\begin{figure}
 \centering
   \includegraphics[width=\linewidth]{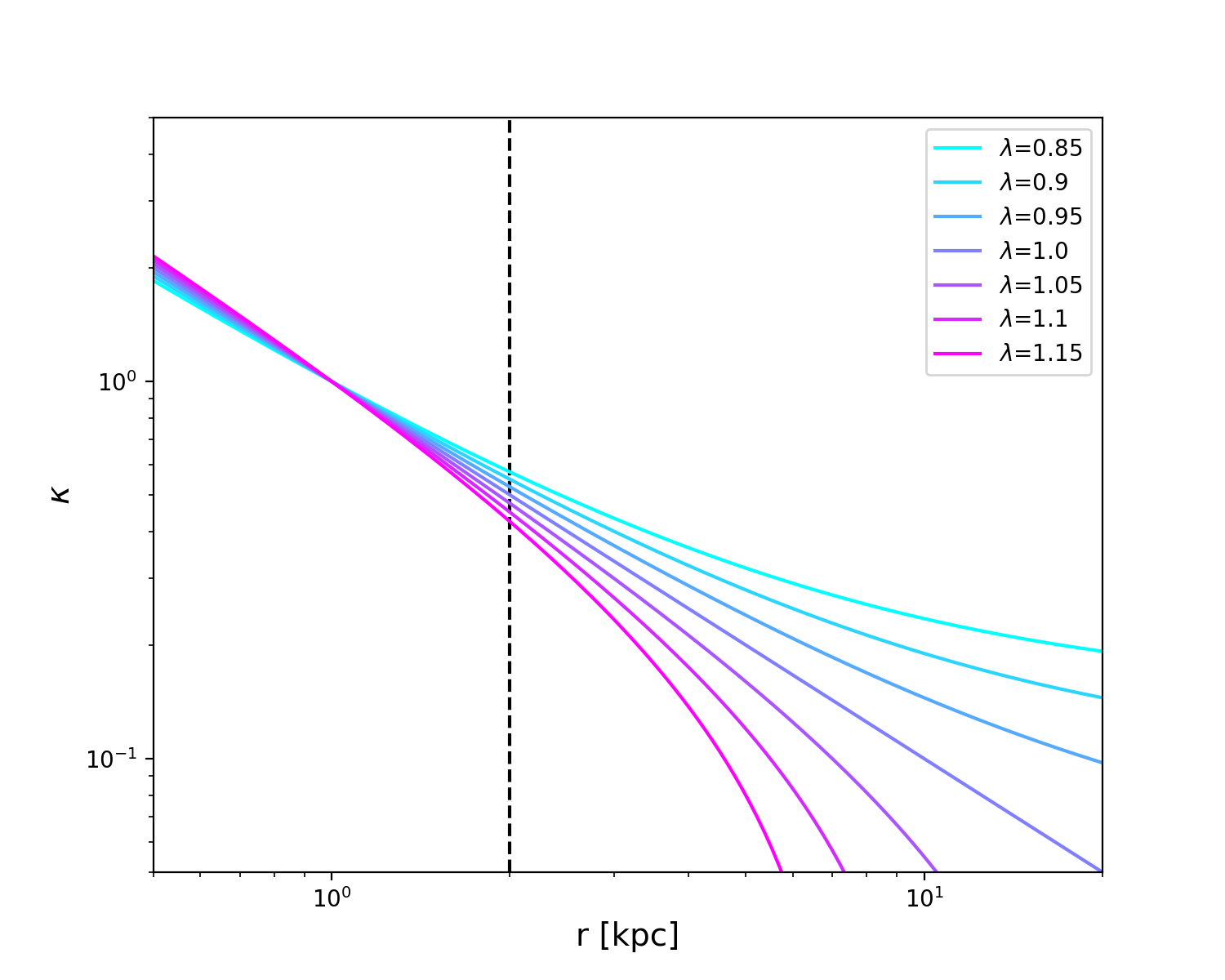}
   \caption{To illustrate the effect of the mass sheet degeneracy, an isothermal power-law profile is transformed by different values of $\lambda$.
           The Einstein radius for this example lies at 2.0 kpc (dashed line). The resulting slope near this radius is quite different 
           for different values of $\lambda$. All of these profiles would be equally well supported by lensing information, 
           but when fitting to a model profile shape, one of these solutions is preferentially selected.
        }
\label{fig:mstlambdafig}
\end{figure}

Meanwhile, $\lambda$ does affect the product of $H_0$ and the time delay:
$H_0\Delta t\rightarrow \lambda H_0\Delta t$, meaning that
the recovered value of $H_0$ will be biased by a factor of $\lambda$. 
For any lens model, the MSD allows for flexibility in the profile shape, since a range of profiles with varying $\lambda$ would all 
reproduce the same observables. In principle, any of those profiles are equally supported by the data, but in practice, only 
one is chosen by the modeling process \citep{Schneider13}. The effect of the MSD on a power-law shape is illustrated in Figure \ref{fig:mstlambdafig} for several
values of $\lambda$.
During modeling, the lens profile is assumed to follow a simple analytical shape, like a power law or a NFW profile. This 
forces $\lambda$ to take on the particular value that makes the profile fit that shape in the region where images are located. 
In this way, a simplifying assumption could impose a value of $\lambda$ (and therefore $H_0$) which is not necessarily the same as 
the true mass distribution, as it has been artificially selected by the model choice. 

It is worth emphasizing that this systematic effect is caused by lensing degeneracies inherent to all lens modeling 
and is not specific to any particular profile. Though this paper will be specifically exploring the effects
with respect to a power-law model, any other profile would also be biased toward the particular value of $\lambda$ which causes the mass 
distribution to most closely match the assumed profile. Even sophisticated methods which use a Bayesian framework
to determine the most likely of multiple different models (such as Autolens, \citet{Night18}) 
will be subject to systematic effects of degeneracies, although the nature of the 
systematic effects will likely be correspondingly more intricate. The exploration of these effects must start with simpler lens models.

The combined effect of the MSD and simplifying assumptions about the density profile shape has recently been analyzed by \citet{Xu16}. The authors extracted galaxies from the Illustris simulation along different lines of sight and looked at their lens profiles. They calculated the $\lambda$ necessary to transform each profile into a straight power-law shape (with slope $s_{\lambda}$) near the image radius and assumed that this would be the value of the multiplicative bias on $h$ a lens modeler would recover when fitting the system with a power-law profile. They found that the mean deviation of $\lambda$ from unity can be as large as 20-50\% with a scatter of 10-30\% (rms). Even limiting their sample to the galaxies which recover a slope near isothermal resulted in a systematic deviation $\sim 5\%$ with a scatter of 10\%, implying that the power-law assumption introduces significant bias in the recovery of $H_0$. More recently, \citet{Tagore18} performed a similar analysis using galaxies from the EAGLE simulation, with similar results. \citet{Tagore18} continued their analysis by supplementing the lens systems with aperture velocity dispersion information. After using a joint model analysis and omitting lenses with poor $\chi^2$, they found that double image lenses were still biased at the 7\% level. Quad lenses were less biased, with the cross quads specifically being the least biased at the 1.5\% level (Table 7 and Figure 11 of \citet{Tagore18}). It may yet turn out that the improvement and inclusion of kinematic information, combined with clever selection criteria, can help to mitigate the effects of the MSD. 

Despite this finding, caution is advisable. Both \citet{Xu16} and \citet{Tagore18} extract lens profiles from state-of-the-art simulations, which may not have the resolution to describe the inner radii in sufficient detail. In particular, both studies selected galaxies with $R_E\geq2\epsilon$, where $R_E$ is the Einstein radius and $\epsilon$ is the gravitational softening length of the simulation, and calculated slope and $\lambda$ by using measurements at $0.5R_E$, i.e. as small as $\epsilon$. These findings are dependent on the simulations being well-resolved at just one softening length. This concern is exacerbated by the recent work of \citet{vdBosch18}. By analyzing a simplified case of a dark matter subhalo orbiting a static potential, they found that tidal distruption of subhalos within simulations is predominantly a numerical phenomenon rather than a physical process, implying even cutting-edge simulations may not be fully converged.

Even relatively small discrepancies between the lens model and the actual profile are cause for concern because first-order perturbations to the profile can produce zeroth-order changes in time delays and therefore $H_0$ \citep{Read07}. Additional cause for alarm has recently been shown by \citet{Kochanek19} and \citet{Blum20}, who demonstrated, using different means, that  lens models with oversimplified or wrong assumptions can lead to high precision, but inaccurate determinations of $H_0$. \citet{Kochanek19} concluded that $H_0$ cannot be more than $\sim10\%$ accurate, despite claims of higher precision.

The goal of this paper is to quantify the bias and spread in the recovery of $H_0$, both with and without the inclusion of stellar kinematic constraints. Rather than drawing lens profiles from simulations, we will create lenses from two-component analytical profiles, constructed to represent both baryons and dark matter. Because of this, the true profile shape is well-known beforehand.

This investigation is a controlled study, with the intention being to test the effects of model assumptions on relatively simple profiles 
rather than attempting to perfectly mimic real galaxies. Real galaxy profiles do differ from a power law (or any assumed model), but the  
deviation from a particular model is dependent on the galaxy. While a particular model profile might provide a good description
for a population of galaxies, individual galaxies vary, and will deviate from the model by varying amounts. 
Simulated halos have significant variance in their shape parameters \citep{Navarro10}, and still lack self-similarity even when
baryons are included \citep{Chua17}. 
Galaxy profiles may be well-approximated by power laws at the 10\% level \citep{Kochanek19}, which has been adequate for galaxy-formation science, but
is a large amount of deviation when compared to the 1\% $H_0$ goal.
Our synthetic lenses (Figure \ref{fig:FourModelProfile}) are consistent with real galaxies, but their deviation from a power law is in a known and  
controlled way, rather than the random, unpredictable deviation of real galaxies. 
This serves as a starting point to test the effects of the power-law assumption.

From these lenses, quads will be created. The image positions and time delays will then be fit using a simple power-law ellipse+shear model, a common model for real systems. We will then compare the resulting slope and $H_0$ with the expected value of $s_{\lambda}$ and $\lambda$ predicted by \citet{Xu16}. Such agreement, which the authors assumed, would mean that it is possible to calculate the bias given the profile shape from simulations, 
while disagreement would mean that the MSD manifests in a more complicated way which is less straightforward to calculate.

In practice, stellar kinematic information can be used to provide an absolute measure of mass, breaking the mass-sheet degeneracy \citep{Suyu14,Barnabe07}.
This extra information is hypothesized to reduce the bias and the spread of $H_0$.
We will explore this hypothesis in two ways. In Section \ref{ssec:kinematics}, we test the effects of constraining the slope
in the parameter recovery. This emulates the additional constraint of kinematics through the inclusion of external information about the mass profile.
In Section \ref{ssec:jeans}, we calculate the integrated velocity dispersion using a spherical Jeans approximation. We compare the velocity dispersion
for the actual profile with what one would find if the profile were the power-law recovered from the lensing information. A comparison of these values
allows us to diagnose whether or not kinematic information can correctly break the MSD when the mass model is slightly oversimplified.

Throughout this paper we will use $h=H_0/100$ km s$^{-1} \text{Mpc}^{-1}$. Lenses are constructed with $h=0.7$.

\section{Preliminary Tests}\label{sec:prelim}
We will be fitting quads using the \texttt{lensmodel} application \citep{Keeton01b}. The application inputs observational constraints combined with a choice of parametric model, then fits the system using the $\chi^2$ calculated by comparing the modeled images to the observed constraints. This application has been used to model strong lens systems in a variety of studies (see \citet{Lefor13} and references therein). Though \texttt{lensmodel} is capable of using image fluxes and extended images, 
we will simply evaluate $\chi^2$ using image positions and time delays as our observable quantities, assuming optimistic observational 
uncertainties of 0.003 arcseconds in spatial resolution and 0.1 days in time delay measurements. 
This spatial resolution is too precise for optical telescopes, but is feasible using VLBI measurements in the radio, which is currently being
done for lenses in the strong lensing at high angular resolution program (SHARP, \citet{Spingola18}). 
We use this uncertainty in the spirit of making the strongest possible constraints on a lens model.
The first step we must take is to 
confirm that we are able to accurately recover lens parameters from our mock quad images.
We conducted several initial experiments to confirm this.

We wish to adopt a commonly-used analytical model with simplifying assumptions about the
mass distribution of the lens. 
Specifically we choose to fit the lens with a ellipse+shear power-law model, which has 7 parameters: mass normalization,
ellipticity, ellipse position angle, shear, shear angle, core radius, and slope. Since there are 9 observations (6 relative image coordinates
and 3 relative time delays) there are $9-7=2$ degrees of freedom. 

Specific to \texttt{lensmodel}, 
we experimented with the \texttt{alpha} and \texttt{alphapot} models,
which are a power-law mass distribution and lensing potential, respectively.
Our preliminary tests were on several basic lenses, some matching the power-law forms of
the fitting models and some using other profile shapes (namely the two-component Einasto profiles of \citet{Gomer18}).
Limited to a cursory search using only a few quads, in some cases the lens parameters were successfully recovered.
For other quads we were less successful, leading us to several main findings: 
\begin{enumerate}
 \item For some quads, fitting for two parameters in a different order resulted in a better or worse fit.
       This custom-tailored parameter search is only possible for a small number of quads modeled on an individual basis. 
 \item Despite the optimization routine of \texttt{lensmodel}, the recovered slope frequently gets stuck at a local minimum near 
       the initial slope guess. We also occasionally found that restarts of the optimization routine would drastically depart 
       from the nearby minimum and return bad fits.
 \item Lenses created from power-law mass distributions (as opposed to lensing potential) had parameter 
       recoveries which were worsened by pixelation and the finite window of lens construction.
 \item Lenses created from Einasto mass distributions frequently had poor parameter recovery when assumed to be a power law. This is likely
       due to a combination of the numerical effect above and the MSD power-law assumption biasing the recovery of parameters. 
\end{enumerate}

We will have too many quads to model each in a unique way, such as customizing the order in which parameters are fit. Interestingly, this problem is becoming relevant for real lens systems as well, as the number of known systems continues to grow. Since human supervision is not feasible at this scale, automation must be the way forward. For real lens systems, automated fitting procedures such as Autolens \citep{Night18} are already being developed. Our modeling is significantly less sophisticated, but will still require an automated algorithm which tries several different runs in \texttt{lensmodel} to find good fits for each quad in a uniformly controlled way. 

Our fitting procedure is devised specifically to avoid the pitfalls of (1)--(4). Here we define our method explicitly. The procedure is nearly identical to the example in \citet{Keeton01b}, with one extra step. The first run fits the quads with only the mass normalization free to vary, while searching a grid over all values for the position angle and shear angle. All other parameters are held at fiducial values for this first run. The robustness of this process against changes to these fiducial values is detailed in Appendix \ref{sec:conchecks}. Next, a run is executed which uses the best fit result from the previous run as an initialization. This second run allows mass normalization, position angle, and shear angle to vary while searching a grid over values of ellipticity and shear. The third run initializes using the best-fit result of the second run and allows all 7 parameters to vary. This third run implements the ``optimize" routine of \texttt{lensmodel}, which uses the \texttt{amoeba} algorithm available in \citet{NumRec}, restarting several times to ensure that the result robustly returns to the same minimum. The last step is an additional step we have added to make sure the slope recovery does not get stuck at a local minimum. This step restarts the process at the first run, with a different initial value for the slope. Once the process is completed over the desired range of slope initializations, only the single result with the lowest $\chi^2$ is kept. This result is the best fit for this model for a single quad, as the procedure systematically searches over the relevant parameters with a variety of initializations and restarts. To circumvent the problem arising from using mass distributions, from now on we will only construct lenses created via analytical lensing potentials (power law or NFW), fit using a power-law potential via \texttt{alphapot}. This requirement means we can no longer use the Einasto form of the lenses constructed by \citet{Gomer18}. 

\section{Lens Construction} \label{ssec:models}
Now that the numerical effects of the process have been limited to the best of our ability, we are ready to create our set of lenses. The lenses are constructed through the combination of two components: a baryon power-law component (\texttt{alphapot}) and a dark matter NFW component, which is analytically expressible as a lensing potential \citep{GK02,MBM03}.

\begin{equation}\label{eq:sumpot}
   \phi=\phi_{bar}+\phi_{NFW}
\end{equation}
where
\begin{equation}\label{eq:alphapot}
   \phi_{bar}(\xi)=b(r_c^2+\xi^2)^{\alpha/2}
\end{equation}
and
\begin{equation}\label{eq:nfwpot}
   \phi_{NFW}(\xi/r_s)=2\kappa_sr_{s}^2 f(\xi/r_s)
\end{equation}
with
\begin{equation*}\label{eq:h(x)}
f(w)= \begin{cases} 
      \ln^2\frac{w}{2} - \text{arcch}^2\frac{1}{w} & (w<1)\\
      \ln^2\frac{w}{2} + \text{arcccos}^2\frac{1}{w} & (w\ge1)\\
      \end{cases}
\end{equation*}

Ellipticity is introduced through $\xi=(x^2+y^2/q^2)^{1/2}$, where $q$ is the axis ratio of the potential. For the NFW component, $w=\xi/r_s$ such that ellipticity is introduced in a consistent way, where $r_s$ is the scale radius of the NFW profile. In total,  six parameters are required to make a lens: $q$, $b$, $r_c$, $\alpha$, $\kappa_s$, and $r_s$. The core softening radius, $r_c$ is set to 0.3 kpc. Note that the 2D slope of the power-law mass distribution will be equal to $\alpha-2$. Lenses are placed at a redshift of $z_l=0.6$.

\begin{figure}
 \centering
   \includegraphics[width=\linewidth]{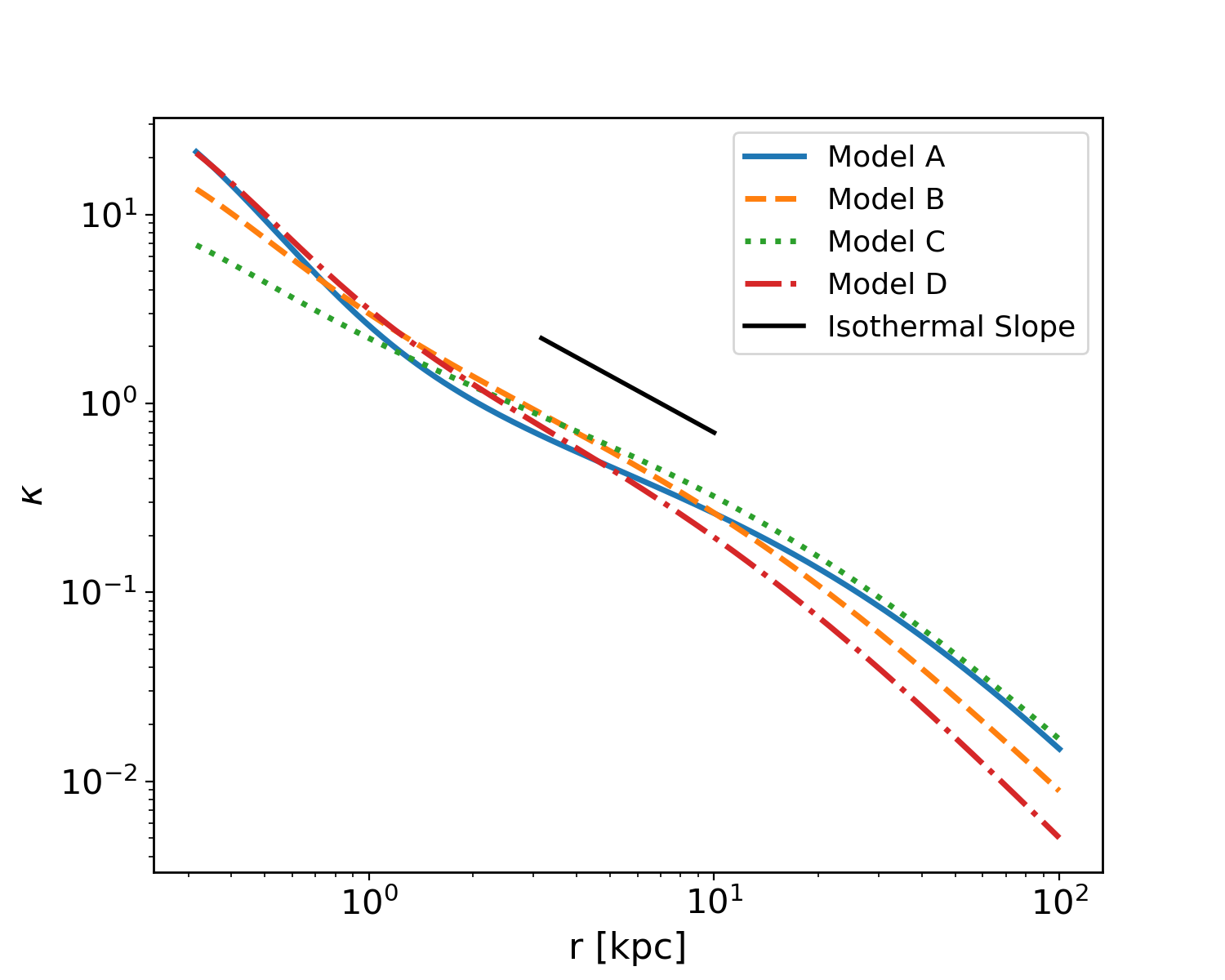}
   \caption{The radial profiles are plotted for the four models. The Einstein radius for each model is set to $\simeq5$ kpc. Note that while models do vary in exact shape, they have approximately isothermal slopes near the image radius. We construct these halos analytically as opposed to extracted from simulations, because the gravitational softening length $\epsilon$ of state-of-the-art simulations is about 0.7 kpc (EAGLE, Illustris). Based on Figure 9 of \citet{Power03}, halos may not be fully resolved until radii 2-3$\epsilon$ outward, meaning simulations cannot reliably detail the nuances of the profile shape interior to about 2 kpc.}
\label{fig:FourModelProfile}
\end{figure}

The cornerstone of the interpretation of \citet{Xu16} is that the value of $\lambda$ calculated from the radial profile will be equivalent to the bias on $h$. To test this, we experiment with a few different sets of values for the parameters which go into making our lenses  (baryon normalization, $b$, and slope, $(\alpha-2)$, as well as dark matter normalization, $\kappa_s$, and scale radius, $r_s$) to create a few different values of $\lambda$ and see how $h$ is recovered in all cases. The choice of profile is somewhat difficult, since many options are physically reasonable. We settle on four different parameter choices of this class of profile, plotted in Figure \ref{fig:FourModelProfile} with their values summarized in Table \ref{table:modelparams}. All of these four models  are comparable to real galaxies.  The Einstein radii, virial radii, and masses are consistent with the EAGLE simulation \citep{Tagore18}. Like real halos, the profiles are nearly isothermal power laws at the Einstein radius (Fig. \ref{fig:FourModelProfile}), and the velocity dispersions are consistent with actual lenses \citep{Suyu10,HC4}. We consider Model D to be  the best analog to a real galaxy due to its slope being slightly steeper than isothermal which is in good agreement with real halos \citep{slacskin3}. Meanwhile, Model A represents the most drastic departure from a power-law model, as evidenced by its visible curvature.

The process to calculate $\lambda$ and $s_{\lambda}$ is quite simple: choose a region near the Einstein radius over which the slope, $s$ is calculated. The magnitude of the mass sheet transformation (MST) necessary to transform the profile into a power law within that chosen region is $\lambda$, while the corresponding new slope after the MST is $s_{\lambda}$ (Equations 4 and 10 of
\citet{Xu16}). 

It should be noted that the same argument can be applied with respect to deflection angle instead of convergence. \citet{Xu16} run the same analysis using the mean $\kappa$ within a radius, referred to as $\overline{\kappa}$, with slope $\overline{s}$. To assume a power law in $\overline{\kappa}$ would be similar to the assumption of a power law with respect to deflection angle. Applying the necessary MST to achieve such a power law would result in a bias $\overline{\lambda}$ and new slope $\overline{s_{\lambda}}$. One advantage of this distinction is that deflection angle is a fundamental lensing property, unlike convergence, which is inferred. Therefore, we must also compare the recovered bias to the analytical prediction of the bias using $\overline{\lambda}$. 

One subtlety here is that the bounds over which the calculation is done are somewhat arbitrary-- \citet{Xu16} and \citet{Tagore18} use $0.5R_E$ and $1.5R_E$ and these are the bounds used for the values calculated in Table \ref{table:modelparams}, but other choices for the bounds are no less valid. Because the slope changes with radius, other choices for the bounds return different values of $s$, $\lambda$ and $s_{\lambda}$. This will be further explored shortly.

For a given model, 100 lenses are created, each producing 1 quad. Each lens is given an axis ratio between 0.85 and 0.99 in the potential, which roughly corresponds to between 0.5 and 0.99 in mass. This range is motivated because values more extreme than $\simeq0.85$ in potential results in mass distribution contours which become ``peanut-shaped'' rather than elliptical. The 100 quads are then fit by the automated process in Section \ref{sec:prelim}, returning values for the 7 parameters, $\chi^2$, and $h$. 

We intentionally choose to fit the lens systems with a model (power law) that does not have the same shape as the density profile (NFW + power law). In real systems, the true mass profile of an individual lens is not directly observable, but some model is assumed based on other studies of a population of galaxies. No individual galaxy will perfectly match the model profile, so this is always the case to some degree. Most studies assume that if the image positions are reproduced, then the lens model sufficiently matches the true mass distribution, but \citet{Schneider13} found that this effect can result in significant bias on $h$. Using a method which explicitly separates the local data-based image constraints from the global model-based assumptions, \citet{Wagner19} showed that image properties can be reproduced without reliance on global assumptions i.e. such that many different global mass distributions are viable. The use of a particular parametric model selects one of these mass distributions over the others, despite not being inherently preferred by the data. Instead, the selection comes from our assumptions about the shape of galaxy profiles in general. Since we can never have perfect knowledge of what the correct profile shape is for a particular galaxy, the effect of our ignorance must be included when seeking to evaluate our ability to fit lensing parameters. 

\begin{sidewaystable}
\centering
\setlength\tabcolsep{4pt}
\begin{tabular}{c | c c c c | c c c c c c c | c c c}
  & \multicolumn{4}{c}{Profile construction parameters}   & \multicolumn{7}{c}{Resulting lens physical attributes}   & \multicolumn{3}{c}{MST values}\\
 \hline
Model  & $\alpha$ & $b [\text{kpc}^{2-\alpha}]$ & $r_{s} [\text{kpc}]$ & $\kappa_{s}$ & $R_E [\text{kpc}]$ & $R_{\text{trans}} [\text{kpc}]$& $\frac{\Sigma_{\text{DM}}}{\Sigma_{\text{total}}}\rvert_{R_E}$ & $M_{200} [\text{M}_{\odot}]$  & $r_{200}[\text{kpc}]$ & $c$ & $\langle \sigma^P \rangle [\text{km/s}] $ & $s$ & $\lambda$ & $s_{\lambda}$\\
A  & 0.30 & 30.0 & 30.0 & 0.12 & 5.1 & 1.7 & 0.81 & $4.5\times10^{12}$ & 271 & 9.0 & 207 & -0.830  & 1.06 & -0.94  \\
B  & 0.60 & 10.07 & 13.33 & 0.206  & 5.7 & 2.1 & 0.67 & $2.8\times10^{12}$& 234 & 17.6 & 262 & -1.03  & 0.925 & -0.895 \\
C  & 0.816 & 3.76 & 22.25 & 0.157  & 4.9 & 1.6 & 0.68 & $5.4\times10^{12}$ & 291 & 13.1 & 332  & -0.815  & 0.904 & -0.702 \\
D  & 0.40 & 22.53 & 10.0 & 0.225  & 5.5 & 2.0 & 0.70 & $1.7\times10^{12}$ & 195 & 19.5 & 225 & -1.14  & 0.960 & -1.05\\
\end{tabular}
\caption{
  The characteristics of the four lens models shown in Figure \ref{fig:FourModelProfile}. First, the parameters used to construct the different lens profiles are presented (outlined in Section \ref{ssec:models}). The next set of numerical values correspond to physical features of each lens profile: the Einstein radius, the radius at which the surface density profile transitions from being dominated by baryons to being dominated by dark matter, the local dark matter fraction at the Einstein radius in 2D projection, the virial mass and virial radius, concentration and the integrated stellar velocity dispersion within an aperture of 1 arcsecond (about 6.7 kpc). Calculations of relevant physical quantities assume a lens redshift of 0.6 with $\Omega_m=0.3$, $\Omega_\Lambda=0.7$, and $h=0.7$. Finally, the values associated with the MST formalism of \citet{Xu16} are calculated: $s$, the slope calculated by using $0.5R_E$ and $1.5R_E$ as the endpoints; $\lambda$, the value required for the mass-sheet transformation to make the profile a power law over the corresponding radii;  and $s_{\lambda}$, the resulting slope of the transformed profile,  which should be the value recovered by the fitting software through the power-law assumption. The values of $s$, $\lambda$, and $s_{\lambda}$ are simply for comparison purposes and are not used in the fitting process.
  }
\label{table:modelparams}
\end{sidewaystable}

\section{Results}\label{sec:results}
We are primarily interested in the statistical results of an array of quad systems. Nonetheless, we have more deeply explored a particular quad from Model D to make certain that the image positions and time delays are properly recovered. We share these fittings in Appendix \ref{sec:singlequad}. Confident that our procedure works, we are ready to discuss the results of the population.

\subsection{Parameter recovery: density slope free to vary}\label{ssec:varyslope}

\begin{figure}
\vspace{-20pt}
 \centering
  \begin{tabular}[c]{cc}
   \begin{subfigure}[c]{0.57\linewidth}
    \includegraphics[width=\linewidth]{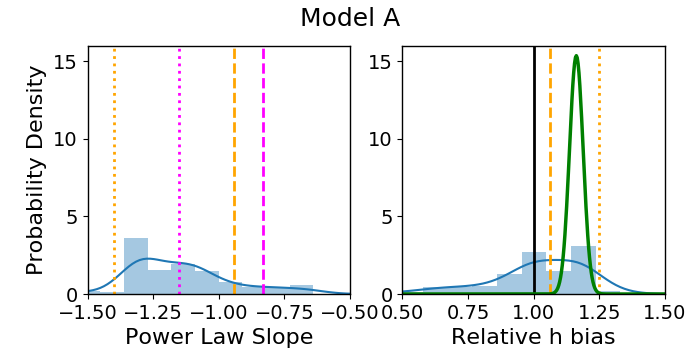}
   \end{subfigure} \\
   \begin{subfigure}[c]{0.57\linewidth}
    \includegraphics[width=\linewidth]{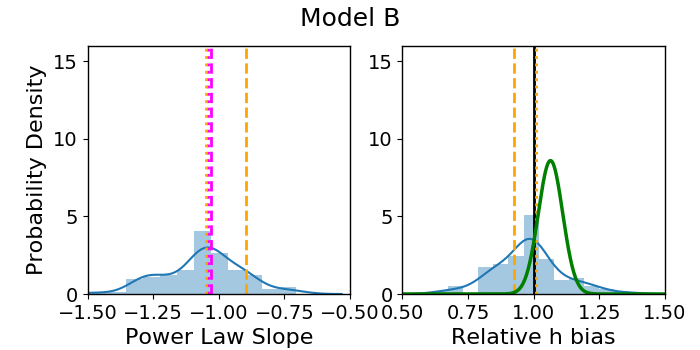}
   \end{subfigure} \\
   \begin{subfigure}[c]{0.57\linewidth}
    \includegraphics[width=\linewidth]{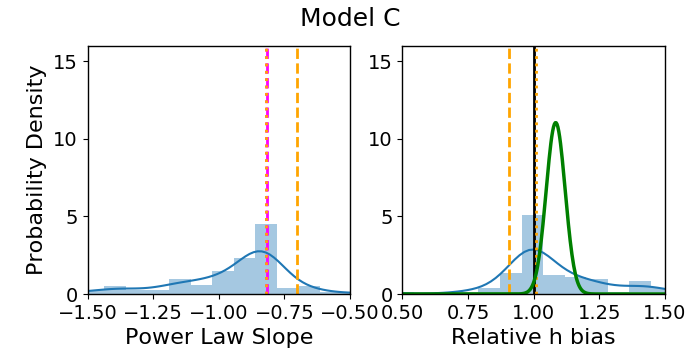}
   \end{subfigure} \\
   \begin{subfigure}[c]{0.57\linewidth}
    \includegraphics[width=\linewidth]{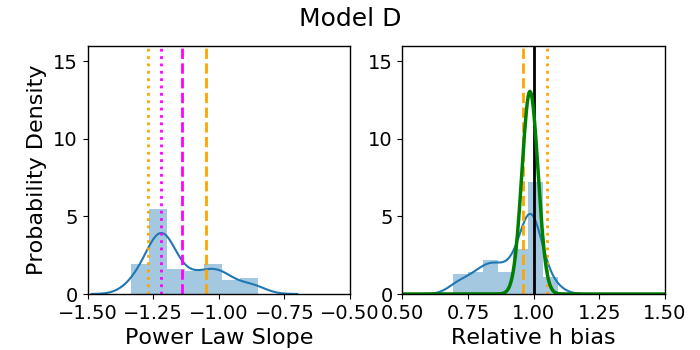}
   \end{subfigure} 
 \end{tabular}
\caption{
   Recovered distributions of power-law slope and $h$ (scaled such that 1.0 is unbiased) are presented after 100 quads are fit for each model. The blue curve/histogram represents the distribution of each best-fit $h$ value, with cases with poor $\chi^2/dof<1$ omitted.  In the left panel, the untransformed estimations of slope are given by the magenta lines (dashed=$s$, dotted=$\overline{s}$), while the transformed predictions of recovered slope are marked with orange lines (dashed=$s_{\lambda}$, dotted=$\overline{s_{\lambda}}$). The transformed predicted slopes do no better than the naive untransformed guess at predicting the recovered slope. In the right panel, the green curve represents the result when combining the likelihoods of each quad together into one estimate of $h$, with error estimated through the bootstrap process detailed in Section \ref{ssec:varyslope}. The solid black line reflects an unbiased value of $h$, while the orange dashed (dotted) lines represent the predicted values of $\lambda$ ($\overline{\lambda}$). The scatter on $h$ evaluated through the MLE is $\simeq3\%$, while the median bias ranges from 2\% to 16\% depending on the model (see Table \ref{table:varybias}). The median bias does not appear to be well-described by $\lambda$ or $\overline{\lambda}$, contrary to the expectation of \citet{Xu16,Tagore18}.
       }
\label{fig:xus&h}
\vfill\null
\end{figure}

The most straightforward way to represent the results is to plot a histogram of the best-fit values of slope and $h$, shown in Figure \ref{fig:xus&h} as the blue distribution for each of the four models. Nearly all ($\ge 97\%$) fits returned $\chi^2/dof<1$. The few cases with bad fits are omitted from
the blue distributions, meaning all of the recovered parameters in the figure are within the uncertainties of observations. As an additional test of modeling success, we checked to see if the recovered ellipticities are strongly correlated with the true ellipticities, and find a Pearson coefficient $R\simeq1.0$ after omitting the few cases with $\chi^2/dof>1$. Our lenses have zero input shear, and the recovered shear values are nearly zero. This tells us that not only do the image positions match, but the mass model parameters correspond quite well to their true values. These measures of fitting success are included in Table \ref{table:varybias}.

While useful, the blue histograms in Figure \ref{fig:xus&h} do not fully capture the process of determining a single value of $h$ from many quads. As the number of systems increases, the shape of the blue distribution will stay roughly the same and will not narrow (see Appendix \ref{sec:conchecks}), as it only returns a single value for each quad and does not combine them together in any way. Meanwhile, determinations of $h$ such as those presented by \citet{HC13} and \citet{Tagore18} represent posterior distributions of $h$ from a single system, as well as aggregated together for a composite distribution from a number of systems.

To evaluate this, we use the \texttt{varyh} function in \texttt{lensmodel} to calculate the $\chi^2$ for a range of $h$ values near the best-fit value, marginalized over the other fitting parameters. \footnote{ 
     This is not strictly equivalent to the posterior distribution one would obtain from an MCMC sampling, but is meant to approximate the distribution. See Section \ref{ssec:limits}.}
We then calculate a likelihood for each quad and combine the likelihoods together to evaluate the $h$ corresponding to the maximum likelihood estimation (MLE).  To quantify the variance of this estimator, we bootstrap the distribution using subsets of 50 quads and evaluate 2000 realizations. The green curve in Figure \ref{fig:xus&h} (right panel) represents a Gaussian fit to the resulting distribution. This curve more accurately depicts the resulting bias and scatter one would get from combining 100 systems together into a single determination of $h$. Table \ref{table:varybias} lists these quantities for each model. \footnote{
    Although the MLE values we quote for $h$ do not omit cases with $\chi^2/dof>1$, cases with poor $\chi^2$ are downweighted in the estimation because likelihoods are calculated from $\chi^2$. We found that when cases with $\chi^2/dof>1$ were omitted, the recovered value of $h$ did not substantially change. (Negligible change for the values in Table \ref{table:varybias} and $\lesssim1\sigma$ for the values in Table \ref{table:slopecalc})}

Across the four models, the distribution of the best-fit $h$ values (blue histogram Fig. \ref{fig:xus&h}) has a scatter $\gtrsim10\%$. As anticipated, combining the fits using the MLE determination of $h$ (green Gaussian) has a much narrower scatter, $\sim3-4\%$.

{\renewcommand{\arraystretch}{1.5}
\begin{table}
\centering
\setlength\tabcolsep{5pt}
\begin{tabular}{c c c c c c c}
\multicolumn{7}{c}{Parameter recovery: density slope free to vary}\\
\hline
Model & $\lambda$ & $\overline{\lambda}$ & MLE $h$ & Mean $\gamma$ & $R_{ell}$ & $f_{\chi^2/dof}<1$\\

A & 1.06 & 1.25 & $1.162\pm0.026$ & 0.011 & 1.00  & 0.99 \\ 
B & 0.925 & 1.01 & $1.064\pm0.046$ & 0.007 & 0.99  & 0.99  \\
C & 0.904 & 1.01 & $1.084\pm0.036$ & 0.008 & 1.00  & 1.00  \\ 
D & 0.960 & 1.05 & $0.986\pm0.031$ & 0.005 & 1.00  & 0.97 \\
\end{tabular}
\caption{The results are presented for the recovery of $h$ when the slope is free to vary in the fitting process. The distribution of $h$ values relative to the correct value, recovered from the MLE, are presented with $1\sigma$ uncertainties. A value of 1.0 corresponds to an unbiased recovery of $h$ while for example a value of 0.986, as in Model D, corresponds to a 1.4\% bias downward. This should be compared to $\lambda$ ($\overline{\lambda}$), which \citet{Xu16,Tagore18} assumed would be the bias in the recovery of $h$ based on the argument that the convergence (deflection) will be transformed into a linear slope over the region between $0.5R_E$ and $1.5R_E$ via the mass-sheet degeneracy. One of our main findings is that the distribution of $h$ does not seem to be related to $\lambda$ or $\overline{\lambda}$, indicating this estimate of bias is not accurate. Also presented are the average shear from the fits, which is near the correct value of zero for all four models, and two measures of goodness of fit: $R_{ell}$ represents the Pearson correlation between the recovered value of ellipticity and the true value, with a strong correlation meaning that the parameter is recovered well in most cases, while $f_{\chi^2/dof}<1$ represents the fraction of systems which were successfully fit within the uncertainties of real observations. Cases with poor fits are heavily downweighted through the MLE process and are explicitly omitted when determining  the mean $\gamma$ and $R_{ell}$.
  }
\label{table:varybias}
\end{table}}

Our main result of this section is that the median recovered values for slope and $h$ do not consistently match the predicted values corresponding to the MST anticipated by \citet{Xu16} (orange dashed and dotted lines, Fig. \ref{fig:xus&h}). For $h$ in particular (right panels), the predicted bias of $\lambda$ should be compared with the MLE result (green Gaussian), and is inconsistent for all but Model D. For the other three models, the prediction misses the mark by 10\%-18\%. In some cases, $\lambda$ underpredicts the magnitude of the bias while in others it overpredicts the magnitude. In both Models B and C, the direction of the bias is incorrectly predicted, failing to even outperform the naive assumption that $h$ will be unbiased (solid black line). Even in Model D, where $\lambda$ is consistent with the MLE result, it misses the mean by about 2\%, which is a significant problem when one considers the 1\% goal.

The capacity of $s_{\lambda}$ to match the recovered value of slope is no more successful. We did not calculate the MLE with respect to slope as our main quantity of interest is $h$. Additionally, to do so would be to assume all quads come from the same profile, which is not true in general. We can only compare to the blue distribution of best fit values. In all but Model A, $s_{\lambda}$ (orange dashed line) makes a worse prediction than the untransformed slope (magenta dashed line).

The prediction coming from a power law with respect to deflection fares only slightly better, as represented by dotted lines in Figure \ref{fig:xus&h}. The $h$ prediction ($\overline{\lambda}$, orange dotted line; right panels) also fails to match the recovered value of the MLE, missing by $\sim1\sigma$ in Models B and D, $\sim2\sigma$ in Model C, and $\sim3.5\sigma$ for Model A (Table \ref{table:varybias}). The transformed slope, $\overline{s_{\lambda}}$ (orange dotted line, left panels) does not fare any better than the untransformed slope, $\overline{s}$ (magenta dotted line) at predicting the recovered slope of the model.

One manifestation of the MSD is a relationship between the steepness of a lens profile and the estimate for $h$. This is subtle, but present in Figure \ref{fig:xus&h}, where the general shapes of the blue distributions for slope and $h$ are more or less mirrored, with steeper profiles resulting in a higher $h$. We will explore this effect more thoroughly in the next section.


\subsection{Parameter recovery: fixed slope}\label{ssec:heldslope} The fully automated method allows the slope to vary when recovering the parameters, but it is also useful to note the results when the slope is fixed. Fixing the slope at a particular value is an act of utilizing additional information which breaks the mass sheet degeneracy. In the context of real lenses, this information can come from the inclusion of stellar kinematics, which probe the mass at radii near the images. When combined with lensing mass estimates, constraints are effectively placed on the profile slope. A truly complete analysis of this effect would be to include a model for stellar kinematics and simultaneously fit velocity dispersion data with the image positions to recover lens parameters. This is beyond the scope of this paper, but we will explore the intricacies of velocity dispersion data more directly in Section \ref{ssec:jeans}. We are still interested in the general effect that arises from knowing information about the slope, and fixing the slope at a particular value approximates the effect.

The question then becomes what value to fix the slope to. Is the ``correct'' value the one which the true mass distribution follows ($s$), the one which corresponds to the slope after the MST molds the profile into a power law ($s_{\lambda}$), or some other slope? An additional complication is that the value for each is dependent on the bounds over which the slope is calculated. Which of these values, if any, will result in zero bias on $h$ recovery? To explore this question, allow us to focus on Model D; we will return to the other models later in this section.

We ran a similar test as the ones before, with 100 realizations of the Model D profile, but this time constraining the slope to be -1.1. This value is chosen because it is close to both $s$ (-1.14) and $s_\lambda$ (-1.05) one would calculate using $0.5R_E$ and $1.5R_E$ as the bounds. Fewer quads are fit with acceptable $\chi^2/dof$ (69/100) but the correlation between model ellipticity and true ellipticity is still very strong. Again combining the fits together into an MLE determination of $h$, the recovered value of $h$ is now considerably biased (-11.5\%, Table \ref{table:slopecalc}). Since slope and $h$ are strongly linked, we interpret this result to mean that the value of slope used here is not the value which would result in zero bias on $h$. There must exist some value of slope which results in an unbiased $h$, but lensing degeneracies have manifested through the modeling process in some way causing this value to be different from what we anticipated. This prompts us to consider the value of the slope more carefully.

\begin{figure*}
 \hspace{-0.5in}
  \begin{tabular}[c c]{cc}
    \setlength{\tabcolsep}{1pt}
    \hspace{20pt} Model A &  Model B \\
   \begin{subfigure}[c]{0.56\textwidth}
    \includegraphics[width=\textwidth]{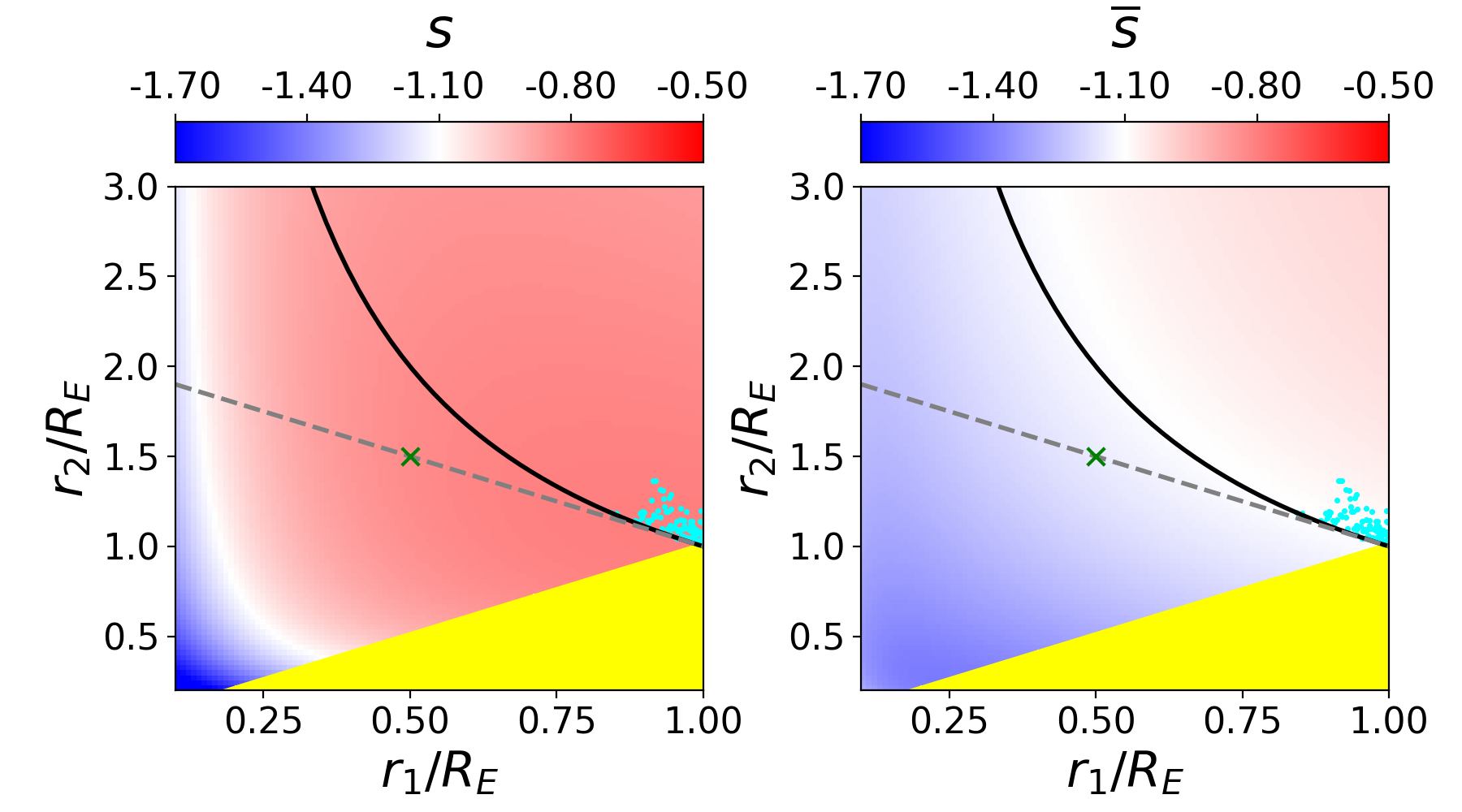}
   \end{subfigure} & \hspace{-15pt}
   \begin{subfigure}[c]{0.56\textwidth}
    \includegraphics[width=\textwidth]{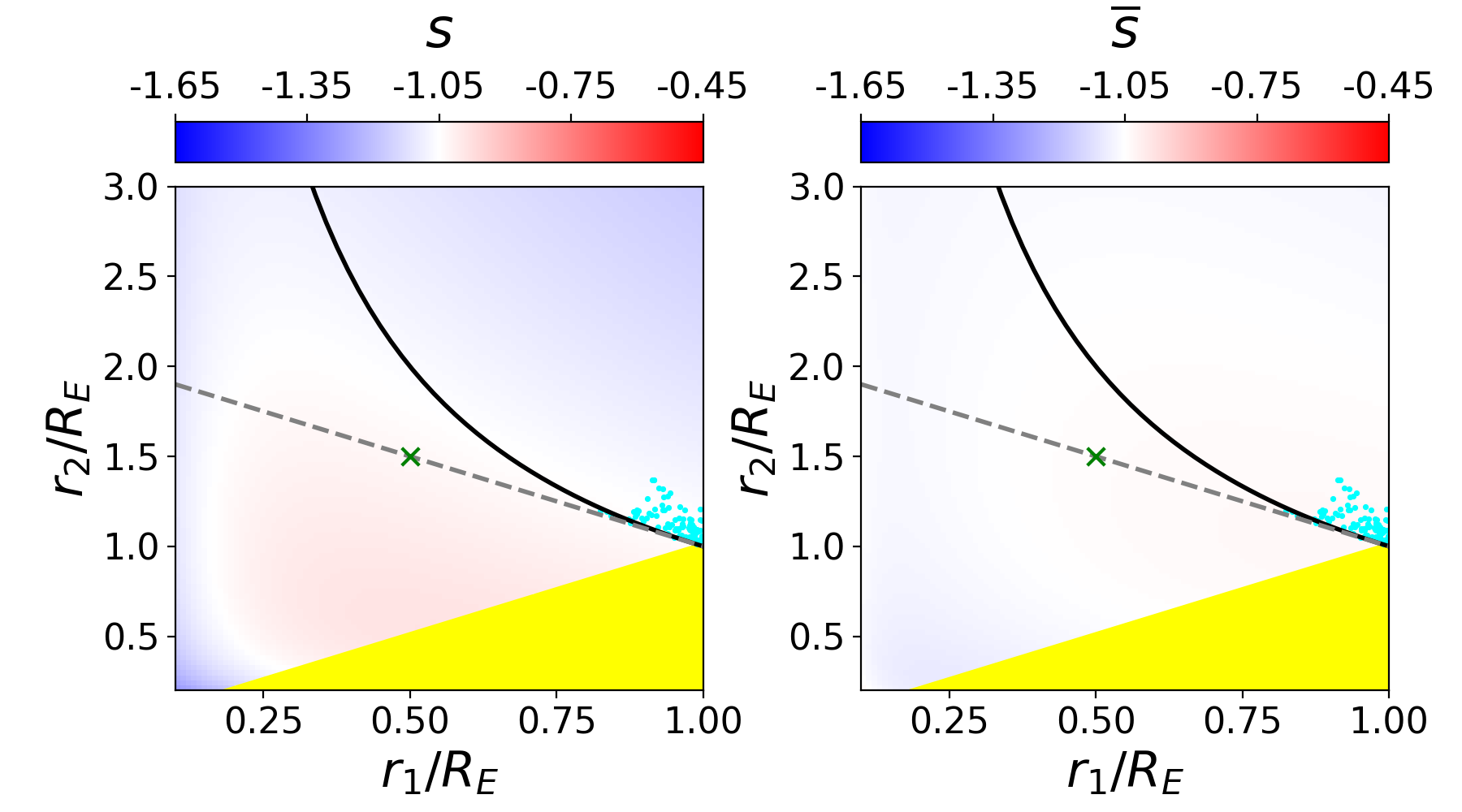}
   \end{subfigure} \\
    \hspace{20pt} Model C & Model D \\
   \begin{subfigure}[c]{0.56\textwidth}
    \includegraphics[width=\textwidth]{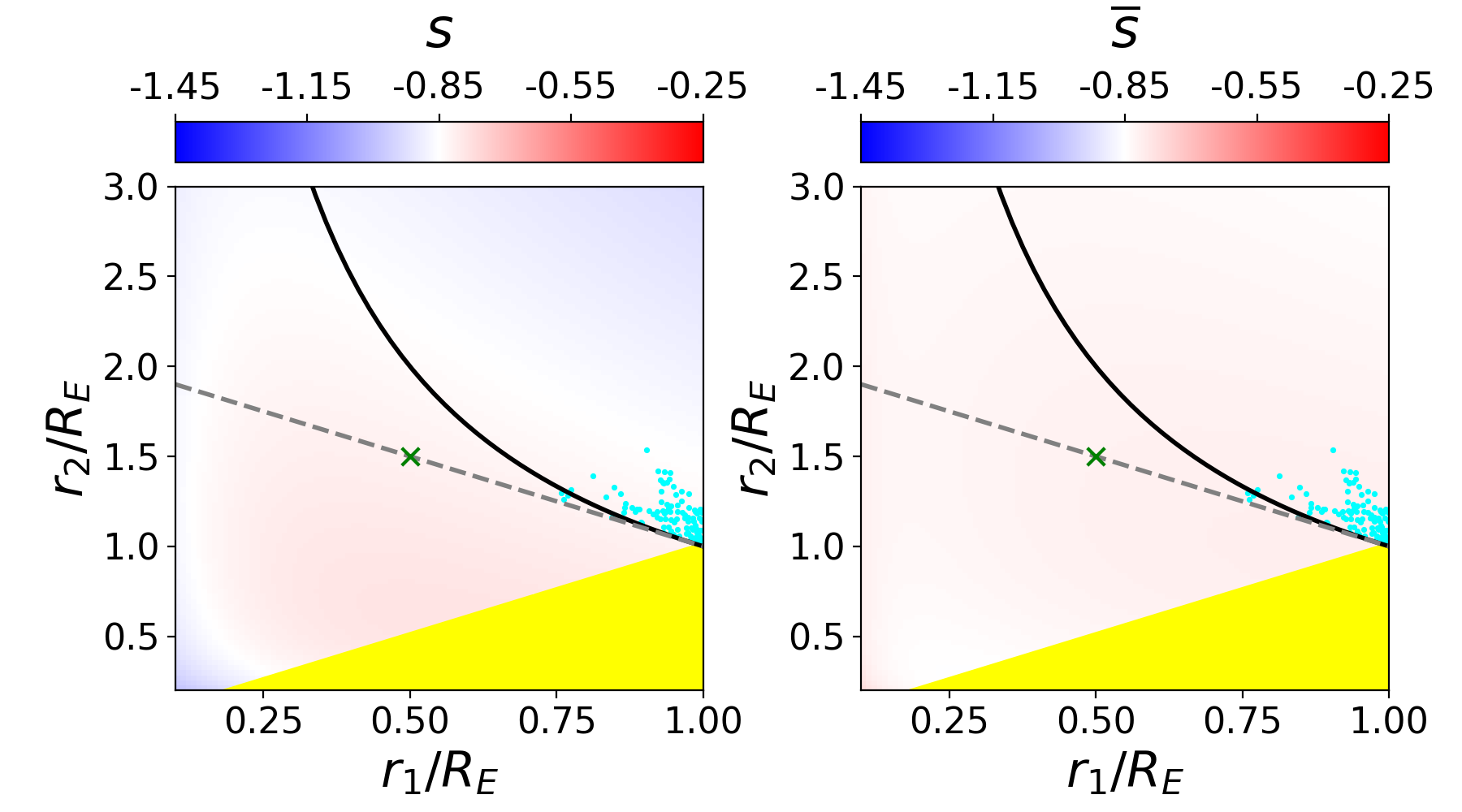}
   \end{subfigure} & \hspace{-20pt}
   \begin{subfigure}[c]{0.56\textwidth}
    \includegraphics[width=\textwidth]{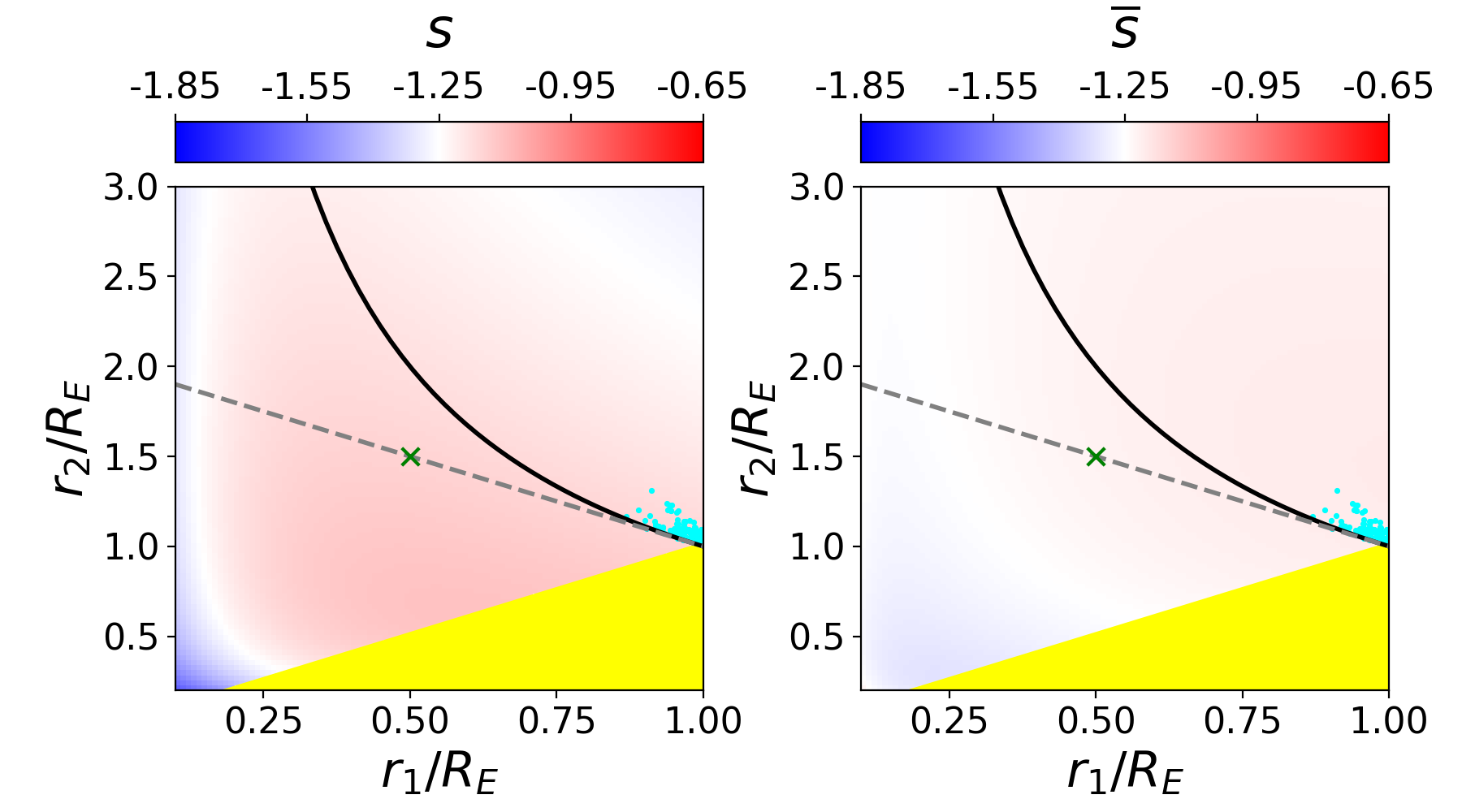}
   \end{subfigure} 
  \end{tabular}
\caption[The dependence of the resulting slope on the region over which it is calculated]{
    The choice of region over which the slope is calculated impacts the measured slope, $s$ (left panels) or $\overline{s}$ (right panels). For each of the four models, vertical and horizontal axes represent the upper- and lower-bound choice for radius within which the slope is calculated, $r_2$ and $r_1$ respectively. Since the upper bound must be greater than the lower, the yellow region is non-physical. Shaded color represents the resulting value of $s$ ($\overline{s}$). Note that the range of values  is different for each and indicated by the color bar in each panel. The colors are set such that the white regions correspond to the slope which results in nearly zero bias (listed for each model in Table \ref{table:slopecalc}). The black solid line represents the choices of bounds which are logarithmically spaced around the Einstein radius, while the gray dashed line indicates bounds which are linearly spaced. \citet{Xu16,Tagore18} chose the bounds indicated by the green ``X.'' For 100 quads for each model, cyan points show the region where images probe. The main feature of note is that most reasonable choices of bounds spanning the Einstein radius result in a $s$ which leads to a biased value of $h$ for all four models. Meanwhile ($\overline{s}$) leads to a less-biased estimate of $h$ for most choices of bounds.
   }
\label{fig:slopecalc}
\vfill\null
\end{figure*}

\begin{figure}
 \centering
   \includegraphics[width=\linewidth]{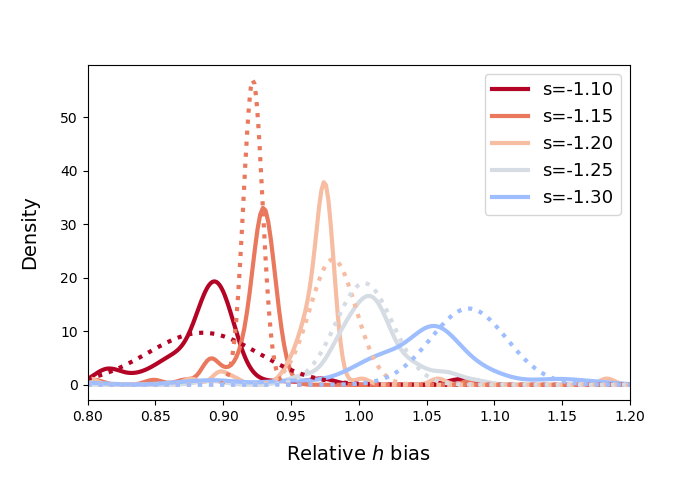}
   \caption{The recovery of $h$ for Model D when the slope is fixed at different values, with colors corresponding to the range of values in Fig. \ref{fig:slopecalc}. For each value at which the slope is held, the distribution of best fit values of $h$ is represented as a solid curve. The dotted Gaussian curves represent the range of values recovered when the fits are combined together and $h$ is calculated through an MLE. Fixing the slope to be steeper results in a higher $h$. The value which corresponds to no bias in $h$ (slope $\simeq-1.25$) is not the same as either the measured slope of the mass distribution or the calculated slope after an MST is applied (both $\simeq-1.1$ for this model). 
   }
\label{fig:ModelDheldslope} 
\vfill\null
\end{figure}

Since the slope of the profile is changing with radius, it is not immediately clear what slope \texttt{lensmodel} should recover. The value of the slope near the Einstein radius is dependent on the choice of the two points used to calculate it. Figure \ref{fig:slopecalc} shows the effect of changing these bounds on the calculated values of slope. The relatively narrow range near the Einstein radius which the images actually span is also depicted (cyan points). Generally, choices which are symmetric about the Einstein radius recover values of $s$ between -1.1 and -1.3 for Model D. It is not obvious which value is the correct one to fix the slope to when recovering parameters in the ``fixed slope'' case. It is therefore prudent to run the ``fixed slope'' test for all values in this range, and see which results in the least-biased value of $h$. The resulting recoveries of $h$ are depicted in Figure \ref{fig:ModelDheldslope}. The MSD is illustrated by a clear trend, where a steeper slope results in a higher value of $h$. The slope value which results in no bias happens to be about -1.25, which is quite different from the value one would calculate using $0.5R_E$ and $1.5R_E$, although similar to the median value in Figure \ref{fig:xus&h} (bottom left). 

We run this same test for all four models, holding the slope fixed at different values. The results are listed in Table \ref{table:slopecalc}. The values of slope which result in the least bias are in bold, while the values with slope closest to $s$ are italicized. In all but Model B, these two values are different. The choice of color scheme for Figure \ref{fig:slopecalc} is now clear, where we have set the white region to the value of the slope which results in no bias. This makes it clear which choices for the bounds on the definition of slope result in the zero-bias value. With the slight exception of Model B, the choice of bounds using $0.5R_E$ and $1.5R_E$ (green cross) is quite removed from the white portions of the figure, indicating this choice of values results in a biased estimation of $h$.

One interesting result which is now apparent is that the value of slope which results in no bias on $h$ corresponds closely to $\overline{s}$, the untransformed slope of $\overline{\kappa}$. This slope also happens to correspond to a density-weighted measure of slope within the Einstein radius, consistently across the four models. Since $\overline{\kappa}$ is an integrated measure of mass, we were curious if this density-weighted measure of slope corresponds to any consistency with respect to the local description of mass, $\kappa$. Exploring this further, we found that the slope of the $\kappa$ profile reaches this value locally at $\simeq0.3R_E$, consistent across the four models. If one could know a priori to fix the slope to this value, one could eliminate the bias on $h$, but the quantity is not directly observable for real lenses.

When the slope is held fixed at a particular value, the scatter of the distribution of $h$ is reduced to $\sim 2\%$, depending on the model and value of slope chosen. This is still too much scatter for a 1\% determination, but it may be that the spread would be further reduced  with additional information coming from extended sources. We are more concerned with the bias, which has a strong relationship with the recovered slope: a shallower slope biases $h$ low, while a steeper slope biases $h$ high. In all cases, the value for the slope which results in minimal bias on $h$ is steeper than both $s$ and $s_{\lambda}$. When the slope is held at values closer to $s$ or $s_{\lambda}$, the recovered value of $h$ ranges from 0--23\% less than it should be, while when it is held at $\overline{s}$, $h$ is recovered without bias. This result in relation to the role of kinematics is discussed in the next section.

{
\begin{table} 
\centering
\setlength\tabcolsep{4pt}
\begin{tabular}{c c c c}
\multicolumn{4}{c}{Model A } \\ 
\hline
Slope & MLE $h$ & $R_{ell}$ & $f_{\chi^2/dof}<1$\\
{\it-0.85} & ${\it0.769\pm 0.023}$ & {\it 0.97} & {\it 0.67} \\
-0.90 & $0.817\pm 0.027$ & 0.98  & 0.70 \\
-0.95 & $0.859\pm 0.013$ & 0.98 & 0.71 \\
-1.00 & $0.907\pm 0.019$ & 0.98  & 0.72 \\
-1.05 & $0.965\pm 0.016$ & 0.97  & 0.58 \\ 
{\bf -1.10 }& ${\bf 1.016\pm 0.023}$ & {\bf 0.99}  & {\bf 0.64} \\ 
-1.15 & $1.078\pm 0.028$ & 0.99  & 0.60 \\ 


\multicolumn{4}{c}{Model B} \\
\hline
Slope & MLE $h$ & $R_{ell}$  & $f_{\chi^2/dof}<1$\\
-1.00 & $0.944\pm0.011$ & 0.99  & 0.60 \\
 \textbf{\emph{-1.05}} &  {\bf \emph{0.975}}$\pm$ {\bf \emph{0.014}} & \textbf{\emph{0.97}} & \textbf{\emph{0.64}} \\
-1.10 & $1.037\pm0.055$ & 0.99  & 0.55 \\
-1.15 & $1.097\pm0.018$ & 0.99 & 0.69 \\

\multicolumn{4}{c}{Model C} \\
\hline
Slope & MLE $h$ & $R_{ell}$  & $f_{\chi^2/dof}<1$\\
{\it -0.80} & ${\it 0.960\pm0.007}$ & {\it 0.98}  & {\it 0.70} \\
{\bf -0.85} & ${\bf 1.023\pm0.008}$ & {\bf 0.98} & {\bf 0.64} \\
-0.90 & $1.067\pm0.019$ & 0.97  & 0.65 \\
-0.95 & $1.130\pm0.018$ & 0.99  & 0.69 \\


\multicolumn{4}{c}{Model D} \\ 
\hline
 Slope & MLE $h$ & $R_{ell}$  & $f_{\chi^2/dof}<1$ \\
-1.10 & $0.885\pm 0.041$ & 0.99 & 0.69 \\
{\it-1.15} & ${\it 0.922\pm 0.007}$ & {\it0.99} & {\it0.68} \\
-1.20 & $0.981\pm 0.017$ & 0.97 & 0.62 \\
{\bf -1.25} & ${\bf 1.004\pm 0.021}$ & {\bf 0.97} & {\bf 0.55} \\
-1.30 & $1.081\pm 0.028$ & 0.99 & 0.67 \\
\end{tabular}
\caption{Resulting recovery of $h$ when power-law slope is fixed at a particular value, scaled such that
     a value of 1.0 corresponds to an unbiased recovery of $h$.
     The value of the slope which results in the least bias is highlighted in bold, while the value
     of the slope which is closest to that of the true mass distribution in the region between $0.5R_E$ and $1.5R_E$, $s$, is italicized. 
     In all but Model B, these slope values do not coincide. The predicted value of slope after an MST, $s_{\lambda}$ (see Table \ref{table:modelparams}),
     is even farther away from the zero-bias values for all but Model A. 
     }
\label{table:slopecalc}
\end{table}}

\section{Discussion}
The motivation of this exploration has been to determine the reliability of the analytical calculation of $\lambda$ using the density profile shape near the Einstein radius as an estimator of $h$. As illustrated in Figure \ref{fig:xus&h}, the distribution of recovered values of slope and $h$ do not correspond to the values predicted using the arguments of \citet{Xu16}. Generally, the distribution of $h$ is no better matched by the predicted bias, $\lambda$, than it is by blindly assuming no bias is present on $h$. Similarly, the mass-sheet-transformed $s_{\lambda}$ is no better than the untransformed slope, $s$, as an indicator of the recovered slope. We see no clear way to predict the bias of $h$ directly from a profile. The intermediate step of creating and fitting realistic mock quads is necessary.

We find this result perplexing, as we found the logic of \citet{Xu16} convincing. We expected that the effect of the mass sheet degeneracy would be to transform the slope into the one which fits the assumed model over the relevant radii. Instead, it appears the degeneracy manifests in a more complicated way. The MSD, or perhaps even some combination of degeneracies, has created minima in the parameter space which do not correspond to the MSD expectation alone.

Beyond this conclusion, our experimentation with constraining the slope has uncovered some interesting results. First, we confirm the relationship between $h$ and slope, where a steeper mass distribution results in a higher value of $h$, a known consequence of the MSD. More interestingly, we find that the slope corresponding to the mass profile near the Einstein radius results in a biased $h$. In other words, even when we give the fitting the ``right answer'' for the density slope it does not result in an unbiased $h$. This merits a discussion of what it actually means when we constrain the slope, what the ``right answer'' really means. For example, if one instead informs the fit with a density-weighted measure of slope, this can result in a unbiased value of $h$. In the end, how does this apply to the physical analog: the inclusion of stellar kinematic information?

\subsection{Kinematic constraints on slope}\label{ssec:kinematics}
When stellar kinematics are included, the profile is probed at a range of radii depending on the spatial resolution of the kinematic information. The exact location of this region is somewhat complicated to evaluate. The constraint itself is an integrated quantity over some aperture radius which is weighed according to its S/N \citep{slacskin1}. Assumptions about the anisotropy of orbits and projection effects introduce additional complications and degeneracies between parameters. The overall effect is to place an integrated constraint on the mass profile over the region, which, when the mass distribution is assumed to be a power law, translates into a constraint on the average slope within the region.

This information is used to break the MSD by constraining the model to have this particular slope. Since we are using a power-law model fit for the lens, this slope constraint is set as the slope for all radii, while in reality the slope changes with radius. This means that the value which the stellar kinematic data recover will depend on the region being probed. We stress that though we are using specific profiles for the true and model profiles, this conclusion applies in general because the true and model profiles will never be identical. Specific to our profiles, we return to Figure \ref{fig:slopecalc} (left panels), which shows the average slope of each profile given the two radii, $r_1$ and $r_2$, used to calculate it. To recover an unbiased value of $h$, the slope has to be measured between the particular radii which result in the white portions of the figure. If stellar kinematics surveys correspond to these regions, the recovered value of $h$ will be reliable, but if they correspond to a red or blue portion, bias will result. 

If the radii probed correspond to a blue region in Figure \ref{fig:slopecalc}, the resulting value of $h$ would be biased high. For example, suppose real halos are more similar to Model B than Model D. The former has a slightly shallower profile, and is nearly isothermal at the image radii. For the Model B profile, when $r_2$ is greater than $2R_E$ ($r>10$ kpc), the determined slope results in a value of $h$ which is biased high. If this were the case in an analysis like the H0LiCOW analysis \citep{HC13}, the result would lead to an overestimation of $H_0$ compared to the CMB \citep{Planck18} and TRGB values \citep{Freedman19}. At present, such a scenario is merely speculation.

It is interesting to note that for all models, there appears to be a region 
near $r_1=0.15R_E$ which results in an unbiased slope. The white region is a nearly vertical strip here, indicating that the value of $r_2$ is less important. The fact that this is consistent across all four models may imply that there may be something special about this determination of slope. In Figure \ref{fig:FourModelProfile} this corresponds to using $r_1=0.75$kpc. It seems feasible by eye that the slope between this radius and, for example, the Einstein radius (5 kpc), reasonably accounts for the baryon component of the profile yet also approximates the slope of the dark matter at farther radii. If $r_1$ were smaller the slope would be too steep at farther radii, but if $r_1$ were larger the slope would be to shallow at inner radii. It appears to be a coincidence, but a consistent one. It may be that if the spatial resolution of stellar dynamics studies can reach this region, the constraint will result in an unbiased value of $h$.

At present, state-of-the-art measurements are insufficient to spectroscopically resolve this region. H0LiCOW \citep{HC4} used 1D spectra from Keck/LRIS to constrain their HE 0435--1223 determination of $H_0$ with a seeing of $0.8''$ (5.3 kpc at $z=0.6$ or $1.1R_E$ in Fig. \ref{fig:slopecalc}). \citet{slacskin4} obtained two-dimensional kinematic data of SLACS lenses using the VLT/VIMOS IFU with a spatial resolution of $0.67''$/pixel (4.4 kpc, $0.9R_E$). To reach $0.15R_E$, a resolution of $0.1''$ is necessary. It could be that this region can be probed without spatially resolving it, since the innermost regions of the galaxy will be brighter and contribute greatly to the S/N of the innermost pixel. Exactly how this enters into the kinematic constraint will depend on how the pixels are weighted, which is outside the scope of this paper.

Unless this region can be reliably probed, the value of $h$ resulting from stellar kinematic constraints will not be unbiased. In fact, if the degeneracy is broken using a different slope, it may introduce more bias than simply not including stellar kinematics at all. For example, Model D returned $h=0.986$ when the slope was free to vary but $h=0.922$ when the slope was held at -1.15, the value of the true slope near the Einstein radius. Pending further investigation into this result, caution may be warranted when interpreting results which include stellar kinematics.

Meanwhile, if stellar kinematics correspond to a measure of the density-weighted slope of the profile (or, equivalently, the local slope at $\simeq0.3R_E$), $h$ may be recovered without bias. This measure of slope corresponds closely to $\overline{s}$ (dotted magenta lines in Figure \ref{fig:xus&h}) and the bold values in Table \ref{table:slopecalc}. Before jumping to conclusions, a more thorough treatment of kinematics must be explored.

\subsection{Inclusion of spherical Jeans kinematics}\label{ssec:jeans}
The act of constraining the slope as a proxy for stellar kinematic information (as we did in Section \ref{ssec:kinematics}) can provide useful insights into this problem, but it would be even better to use the same method as H0LiCOW: to use integrated stellar velocity dispersion to constrain the fitting procedure, breaking the MSD. Since our modeling framework is limited to using only the image positions and time delays, to emulate the full process will require future work. However, we can calculate the integrated stellar velocity dispersion of a given lens and compare them to the dispersion of the power-law model. A comparison of these values can elucidate the findings of the previous section-- if the MSD is correctly broken by the inclusion of integrated lens dynamics, then the model which most closely matches the kinematic information should be the unbiased one. The above finding that the slope constraint biases $h$ predicts that this will not happen. Instead, because the model does not exactly match the true mass distribution, the case which matches the kinematic information will correspond to a biased model.

We calculate the projected velocity dispersion following the framework of \citet{Suyu10}, solving the spherical Jeans 
equation\footnote{A typographical error in \citet{Suyu10} does not square the $\sigma_r$ this equation.}:
\begin{equation}\label{eq:jeans}
 \frac{1}{\rho_*}\frac{d(\rho_* \sigma^2_r)}{dr}+2\frac{\beta_{\text{ani}} \sigma^2_r}{r}=-\frac{GM(r)}{r^2}.
\end{equation} 
The 3D baryonic mass distribution is given by $\rho_*$, while $M(r)$ refers to the total mass, including dark matter. The anisotropy term $\beta_{\text{ani}}=r^2/(r_{\text{ani}}^2+r^2)$, 
parameterized by $r_{\text{ani}}$, encodes the transition from orbits being isotropic in the center to radial at outer radii. 
In general, this anisotropy radius is a fitting parameter in 
stellar modeling, but is set to 4.5 kpc (about 80-90\% $R_E$) in this analysis to serve as a control variable consistent across all lenses. 
For reference, the range for this parameter used in the H0LiCOW analysis has a prior which spans from approximately $0.5R_E$ to $5R_E$ \citep{HC4}.
From this equation, the radial stellar velocity dispersion, $\sigma_r$ can be calculated given a baryon distribution and a total mass distribution. Then, the velocity dispersion
can be weighted according to the light and projected into 2D (Equation 21 of \citet{Suyu10}):
\begin{equation}
 I(R)\sigma_s^2=2\int_R^\infty \left( 1-\beta_{\text{ani}}\frac{R^2}{r^2} \right) \frac{\rho_* \sigma^2_r dr}{\sqrt{r^2-R^2}},
\end{equation} 
where $I(R)$ is the light distribution as a function of 2D radius $R$ and $\sigma_s$ is the projected velocity dispersion. The constraint itself, $\langle \sigma^P \rangle$, is an integrated measure 
of this quantity over a given aperture $\mathcal{A}$. For simplicity, we omit the convolution with seeing included in \citet{Suyu10}. 
\begin{equation}\label{eq:projint}
\langle \sigma^P \rangle^2=\frac{\int_\mathcal{A}I(R)\sigma_s^2 R dR}{\int_\mathcal{A}I(R) R dR}.
\end{equation}

\citet{Suyu10} used a Hernquist profile for the baryons and a power law for total the mass profile, but with this framework in place we can use any model, 
although the Jeans equation may need to be solved numerically. First, we calculate the actual dispersions one would get with our four two-component models. We set
the aperture radius to be 1 arcsecond, which corresponds to about 6.7 kpc. These velocity dispersions are listed in Table \ref{table:modelparams}.

Next, we calculate the dispersions one would get if the total mass were a power law, instead of our two-component profile.
This is calculated the same way, using Eq. \ref{eq:jeans}-\ref{eq:projint}, with the same anisotropy radius and aperture radius, with the only change being that
the total mass, $M(r)$ in Eq. \ref{eq:jeans}, goes as a power law instead of the correct profile. Importantly, we input the same baryon distribution as the actual lens, 
which means that the measurement is done with perfect knowledge of the true light distribution, but assumes slightly incorrectly that the total mass distribution goes simply as a power law. 
With the framework in place we can calculate what the projected velocity dispersion would be if the lens profile were actually a power-law mass distribution.
We will explore several power laws over a range of slopes and normalizations to see whether or not
the power laws which return the correct value of $h$ also match the projected velocity dispersion.
Like the SPEMD model in H0LiCOW \citep{HC4} we implement a power-law fit using lensing information and calculate the corresponding velocity dispersion in same way as
\citet{Suyu10}, but unlike H0LiCOW we do not combine the results together, instead examining the constraints separately.

\begin{figure*}
 \hspace{-20pt}
  \begin{tabular}[c c]{cc}
   \begin{subfigure}[c]{0.51\linewidth}
    \includegraphics[width=\linewidth]{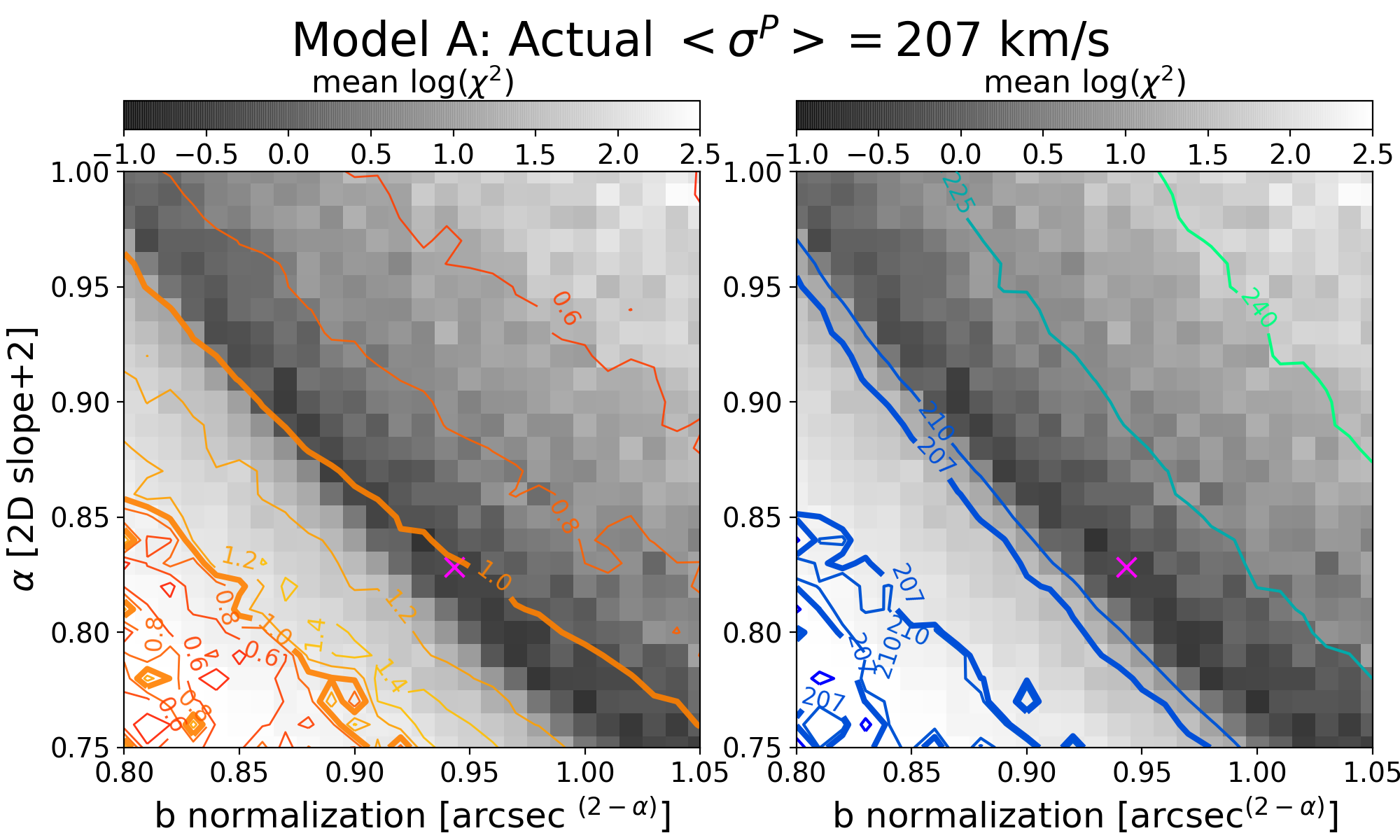}
   \end{subfigure} &
   \begin{subfigure}[c]{0.51\linewidth}
    \includegraphics[width=\linewidth]{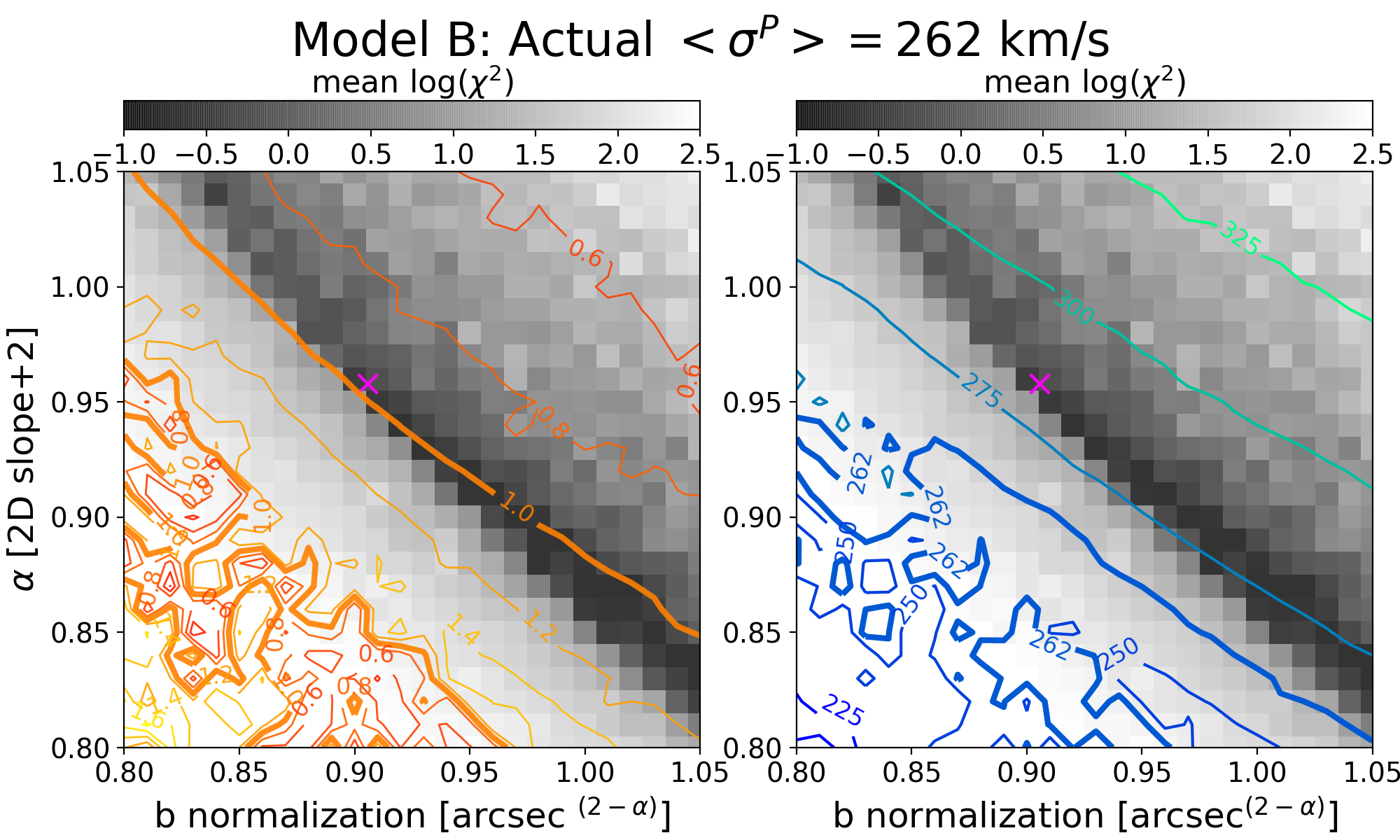}
   \end{subfigure} \\
   \begin{subfigure}[c]{0.51\linewidth}
    \includegraphics[width=\linewidth]{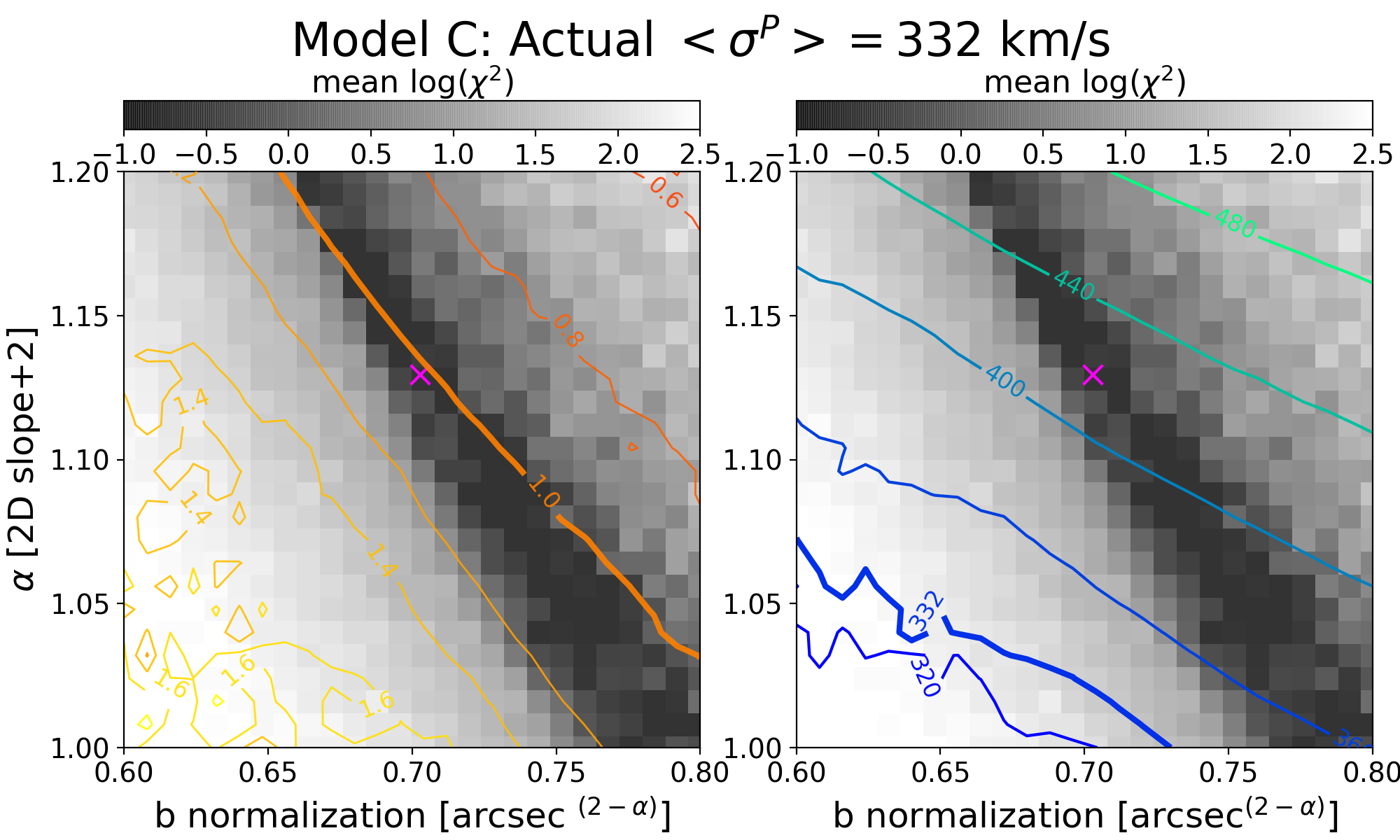}
   \end{subfigure} &
   \begin{subfigure}[c]{0.51\linewidth}
    \includegraphics[width=\linewidth]{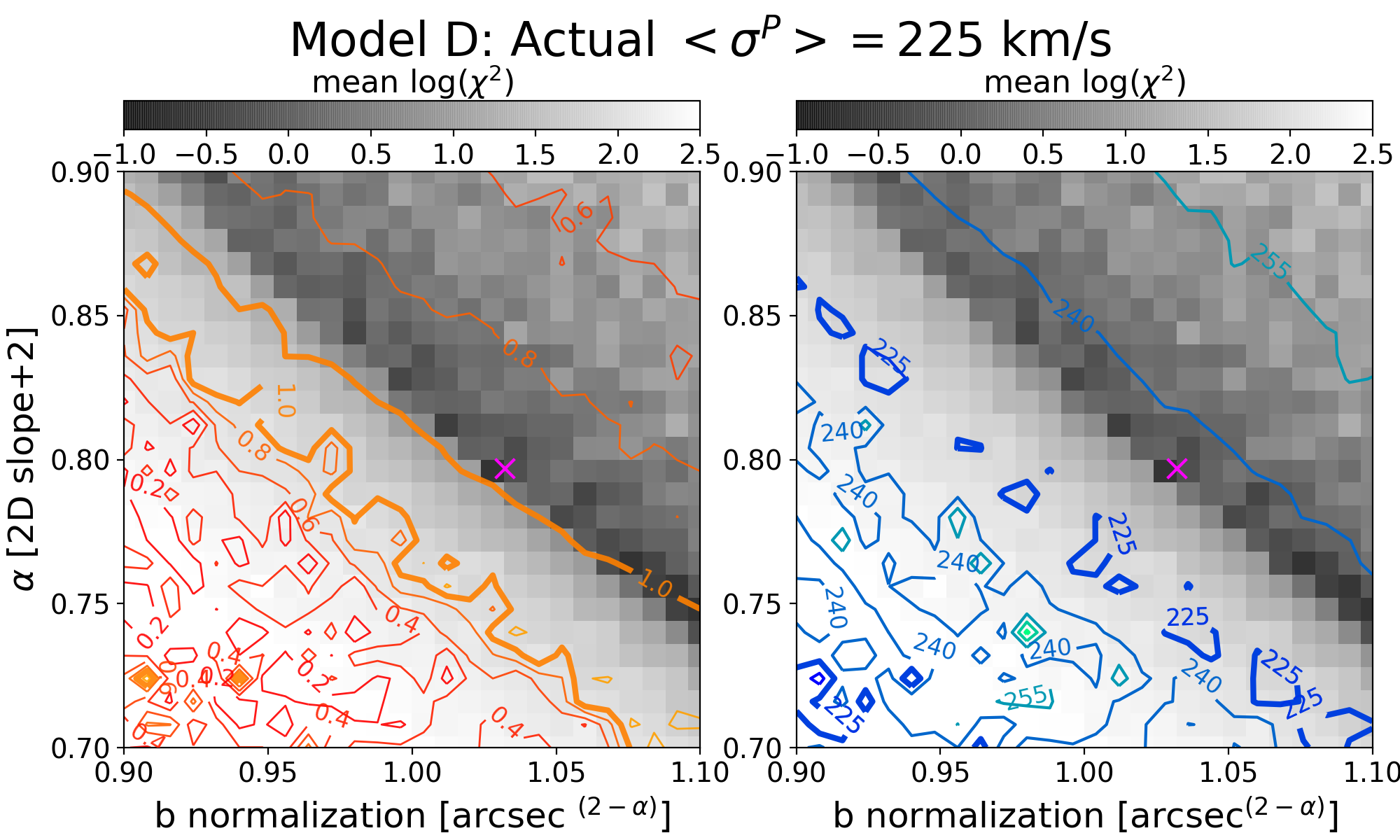}
   \end{subfigure} 
 \end{tabular}
\caption{ For each of the four models, a set of power-law mass distributions is created near the best \texttt{lensmodel} fit from where the slope was allowed to vary (magenta cross). 
      The x- and y-axes represent the two parameters which set the power-law profile: the normalization, b, and $\alpha$, related to slope (see Equation \ref{eq:alphapot}), respectively.
      Each point on the grid corresponds to a power-law mass distribution, for which 50 quads are fit and a $\chi^2$ (grayscale) and $h$ (scaled such that unbiased=1.0, left panel contours) 
      are calculated. On the right panel, the contours show the integrated stellar kinematic constraint (in km/s) calculated for each power law. 
      This value should be compared to the actual $\langle \sigma^P \rangle$ one would observe for each model (thick blue contour) listed above each panel and in Table \ref{table:modelparams}. The dark gray band corresponds to the MSD, 
      where, using lensing information alone, the power law is a decent fit for the model for a range of values. The key feature of note is that, for all four models, the 
      region where the stellar kinematic
      value would match the correct value one would measure (thick blue contour) does not correspond to either the region where the 
      lensing fits have good $\chi^2$ (dark pixels) or where $h$ is unbiased (thick orange contour). To force the fit 
      to conform to the kinematic constraint would pull the fit even farther away from the unbiased answer.
     }
\label{fig:sigmaph}
\end{figure*}

For each of our four models, we explore a set of power-law mass distributions with differing slopes and normalizations ranging near the best 
\texttt{lensmodel} where the slope is free to vary. For each combination of the two power-law parameters, we use \texttt{lensmodel} to calculate the $\chi^2$ for 50 quads and plot the average 
in Figure \ref{fig:sigmaph}. The two panels of the figure show the resulting average $h$ for each combination and the integrated projected stellar velocity dispersion. 
Comparison of these three regions-- where the lensing fits are good (dark gray pixels), where the values of $h$ are unbiased (thick orange contour), and where 
the velocity dispersion measurement corresponds to the correct value (thick blue contour)--provides some interesting conclusions.

First and foremost, the stellar kinematic constraint does not correspond to the regions where the lensing fits are acceptable. For each model, the set of profiles
where the stellar kinematic measurement would match the actual kinematics of the lens has a lower normalization and steeper slope (lower left region in Fig. \ref{fig:sigmaph})
than the lensing result would indicate. This arises because while the light distribution is known exactly, the power-law model is not exactly correct with regard to the
total mass distribution. A Bayesian analysis which combines likelihoods from both lensing and stellar kinematics would pull the best fit downward 
toward this region, driven primarily by the power-law assumption rather than directly by data.

The contours in this region are jagged and unreliable because the this region has poor fits to the lensing information. This directly affects the determination of $h$, but also
indirectly affects the stellar kinematic measurement because the physical conversion scale of kiloparsecs to arcseconds is set by $h$. Because of this, it is difficult to pin down exactly
where the stellar kinematic constraint would place the fit, and also unclear on the exact value of $h$ which would be returned. What is clear is that it would be a bad fit with an unreliable
determination of $h$ which is neither accurate nor robust. In reality, neither the lensing fit nor the kinematic fit is used, but a compromise is sought between the two using a Bayesian 
framework. In this case, the compromise would be between a nearly correct solution and an unreliable solution, a worse result than using lensing information alone.

One further result evident in this figure is that the contours of velocity dispersion run roughly parallel to the dark strip where the model fits the lensing information.
This is interesting because the goal of using stellar kinematic information is to break the MSD and return the unique solution which corresponds to the galaxy profile.
This is not possible if the $\langle \sigma^P \rangle$ contours run parallel to the MSD region because then they are degenerate-- one value of $\langle \sigma^P \rangle$ would 
match all values of $\lambda$ and would not provide unique information. We would be stuck back where we started: with a family of solutions which all match the data. To break the 
degeneracy, the $\langle \sigma^P \rangle$ contours must run at an (ideally perpendicular) angle with respect to the MSD, such that only one unique profile matches both the lensing and kinematic
information. 

To further explore the relationship between these constraints, one can use scaling relations to compare the enclosed mass of a profile within the Einstein radius (which lensing measures)
and the integrated stellar velocity dispersion within an aperture radius. This is detailed for a spherical power-law profile in Appendix \ref{sec:analyticalsigma}.
When the Einstein radius and $\langle \sigma^P \rangle$ aperture radius are similar, the two measurements
are closer to degeneracy-- they are similarly unable to differentiate between a steeper profile and a shallower one, provided the enclosed mass is the same. When the aperture radius is
$0.1R_E$, the contours are nearly perpendicular. The measurement at two different radii provides the information necessary to break the degeneracy, 
supporting the arguments of Section \ref{ssec:kinematics}.

The context of this discussion has been limited to  examination of exact $\langle \sigma^P \rangle$ contours with no accounting for uncertainties. 
In real observations, $\langle \sigma^P \rangle$ is only measured to within about $15-25$ km/s \citep{Suyu10, HC4, ChenHC19}.
With the inclusion of these comparatively large uncertainties, the fit need not be so far down into the lower left regions of Figure \ref{fig:sigmaph} to achieve consistency with the actual 
value for each profile. Instead, it is possible to overlap the lensing fit with the kinematic measurement to within $1\sigma$.
This result would not be informed by a correct breaking of the MSD, but rather happens to be consistent by chance due to the relatively large uncertainties of the stellar kinematics.
The kinematic constraint weights the fit in a direction which has no correspondence with the real lens because the model is misinformed. Perhaps it is fortunate that the 
uncertainties are large so the strength of the weighting is minimal. The logical prediction is that as uncertainties improve in kinematic measurements, 
they will be more heavily weighted and may pull the model parameters farther from where $h$ is unbiased.

\subsection{Subsample selection}\label{ssec:subsample}
\begin{figure}
 \centering
  \begin{tabular}[c]{cc}
   \begin{subfigure}[c]{0.55\linewidth}
    \includegraphics[width=\linewidth]{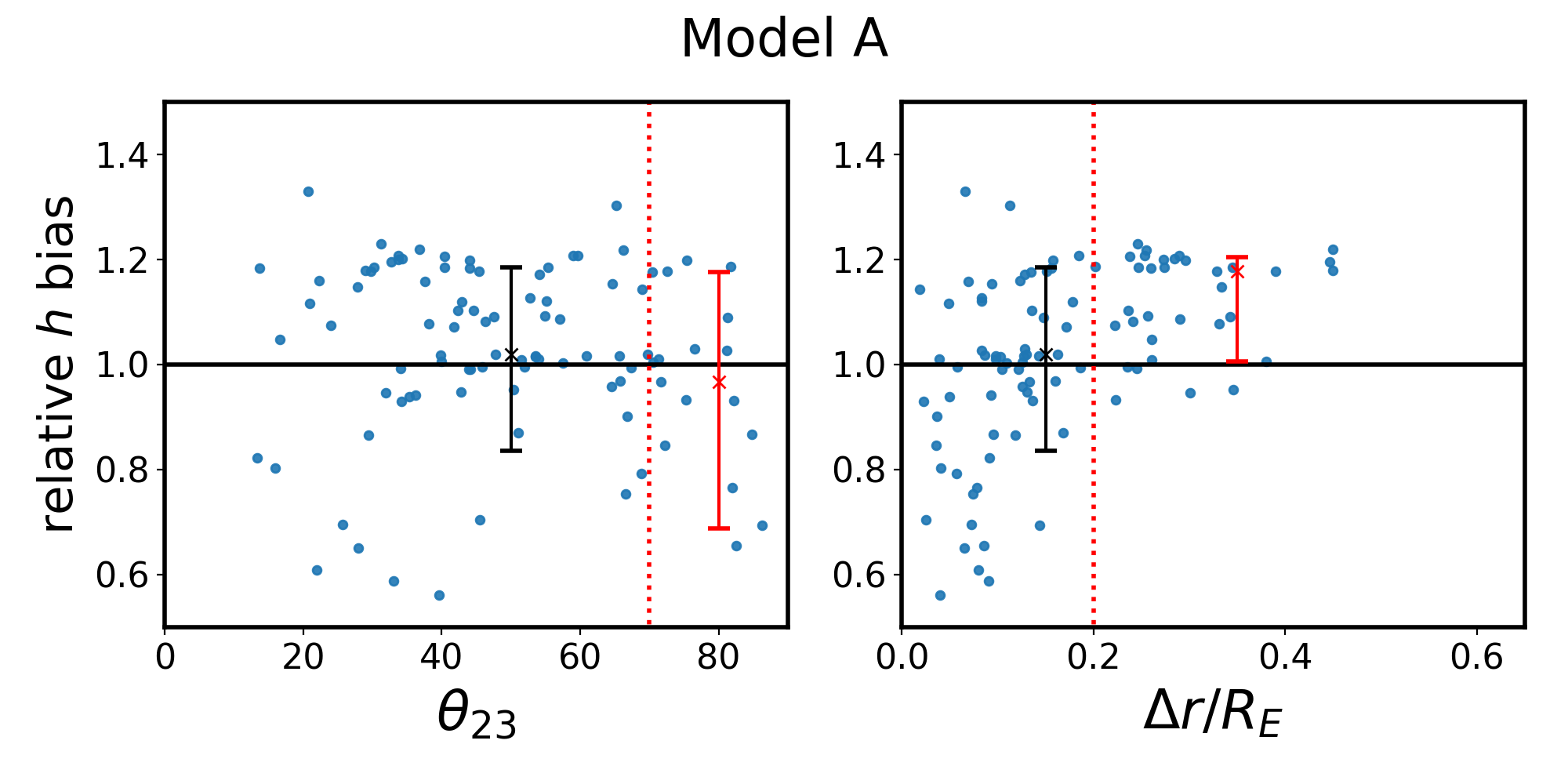}
   \end{subfigure} \\
   \begin{subfigure}[c]{0.55\linewidth}
    \includegraphics[width=\linewidth]{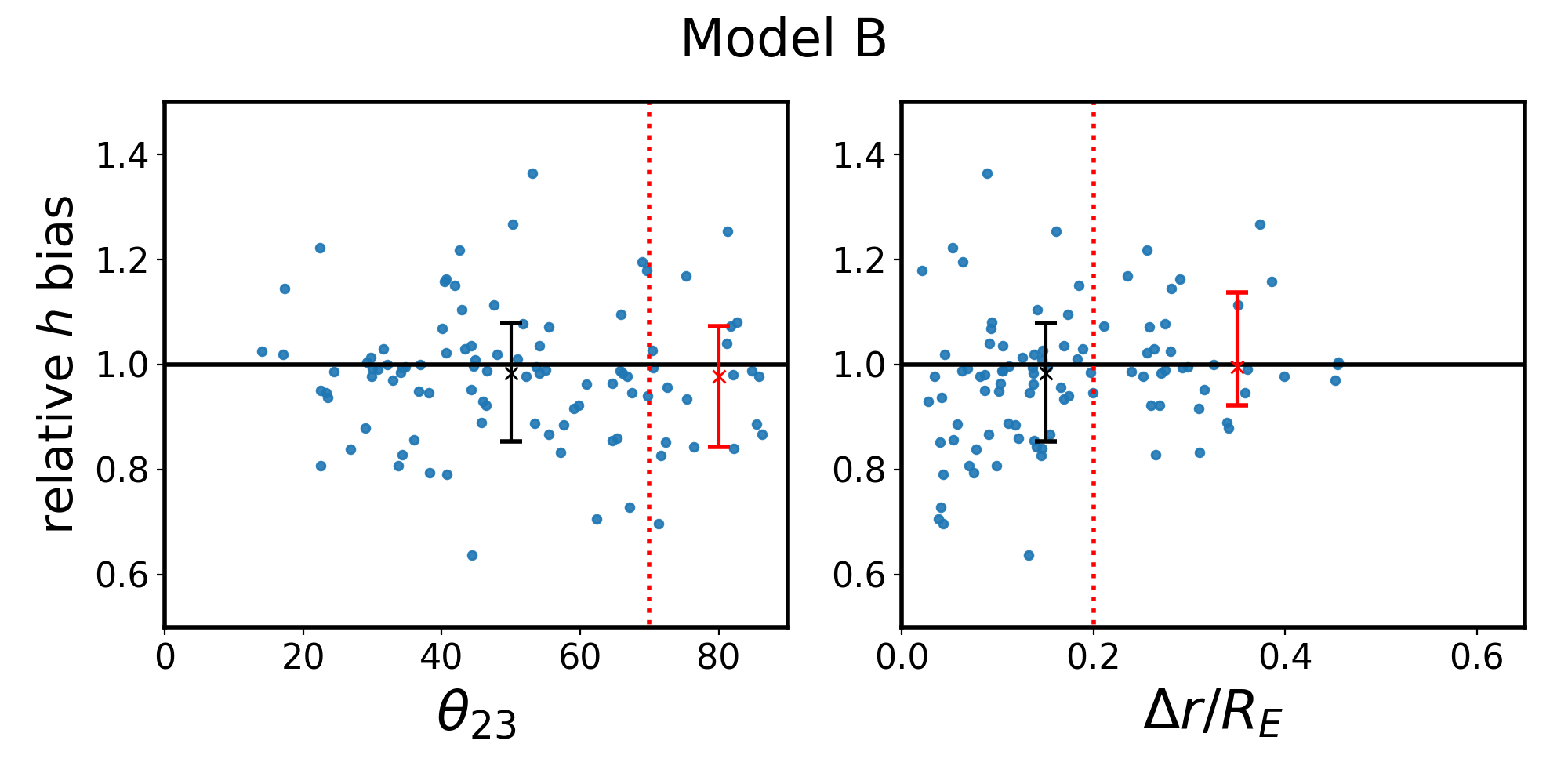}
   \end{subfigure} \\
   \begin{subfigure}[c]{0.55\linewidth}
    \includegraphics[width=\linewidth]{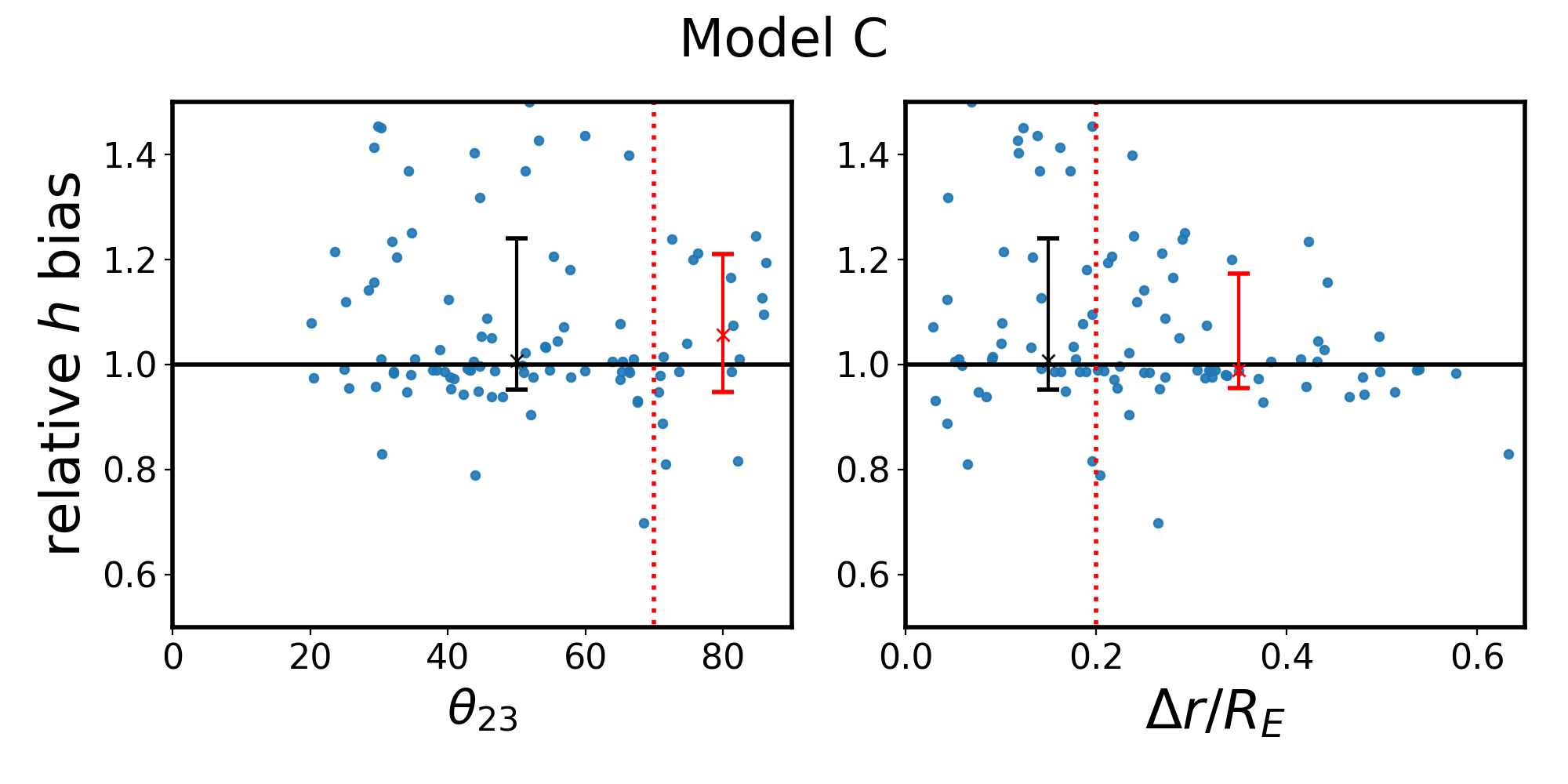}
   \end{subfigure} \\
   \begin{subfigure}[c]{0.55\linewidth}
    \includegraphics[width=\linewidth]{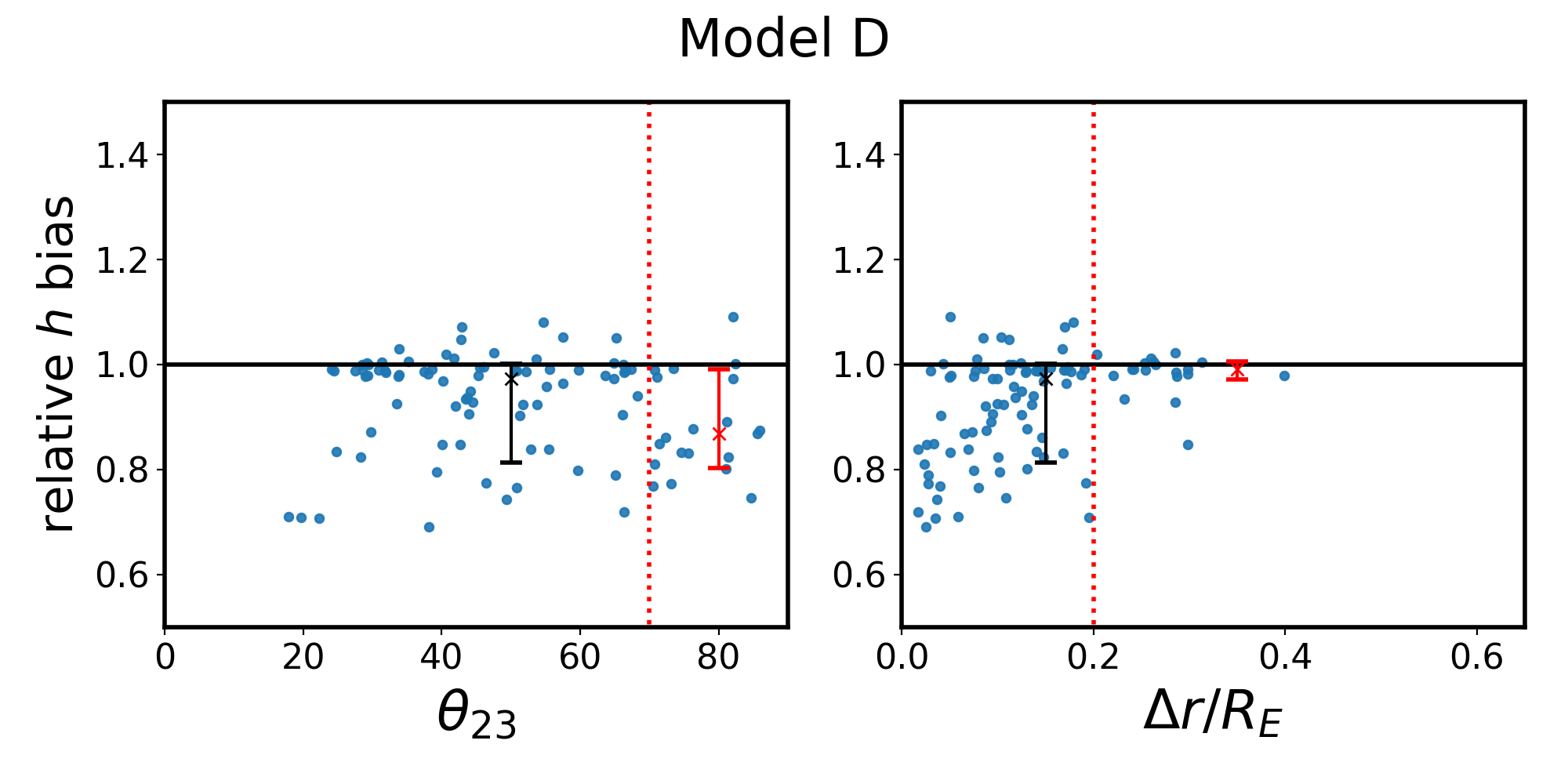}
   \end{subfigure} 
 \end{tabular}
\caption{Scatter plots of the recovered best-fit values of $h$ for each quad against quad orientation (left) and image radial range (right). 
     For each model, the left panel plots $h$ against $\theta_{23}$, the polar angle
     between the second- and third-arriving images, which denotes cusp/core quads (small $\theta_{23}$) from cross quads ($\theta_{23}\simeq90^{\circ}$).
     The black cross with error bars denotes the median and spread of the full set of quads, while the red cross with error bars indicates that of the cross quads: only
     quads with $\theta_{23}>70^{\circ}$ (right of the red dotted line).
     There does not appear to be a significant change in the distribution of $h$ between different quad types.
     The right panel plots $h$ against the radial range over which the images span, $\Delta r/R_E$. Again the black cross with errors indicates the same bias and 
     spread of the whole population, while this time the red cross with error bars specifically refers to the quads with $\Delta r/R_E>0.2$ (again right of the red dotted line). 
     Quads which span a greater range of radii tend to have less scatter in their recovered values of $h$.
     Quads with poor recoveries ($\chi^2/dof>1$) are omitted from these plots. 
     }
\label{fig:scattertest}
\vfill\null
\end{figure}

As a final investigation, we are curious if there exists a subset of quad systems which have distributions of 
$h$ with either less bias or less scatter. To be useful, this selection would need
 to be based on an observable quantity independent of the modeling process. \citet{Tagore18} explored 
the effect of quad configuration (e.g. cusp, fold, and cross) on the recovery of $h$ and found that cross lenses had the least bias.
We adopt the notation of \citet{WW12}, who investigated the angular positions of quad images, wherein the polar image angle between the 
second- and third-arriving images, $\theta_{23}$, serves to represent quad configuration 
(fold and cusp quads have $\theta_{23}\simeq 0$, while cross quads have $\theta_{23}\simeq90^{\circ}$).  
In order to see trends in the MLE determination of $h$ with respect to $\theta_{23}$, it would be necessary to bin the data, which 
would in turn reduce the sample size so low as to make the MLE error estimation unreliable. 
Instead, we simply create scatter plots of the best fit $h$ for each quad versus $\theta_{23}$ (left panels of  Figure \ref{fig:scattertest}).
There does not appear to be a significant reduction in scatter or bias for cross quads as opposed to others.

While we did not find dependence on the type of quad, we also explored dependence on the radial positions of quad images.
The right panels of Figure \ref{fig:scattertest} show that quads which have images over a 
larger range of radii $\Delta r/R_{E} > 0.2$ have less scatter in their recoveries of $h$
than those with a more confined range of image radii. 
To quantify this, we calculated the distributions of $h$ if one selects only 
quads with $\Delta r/R_E>0.2$, to be compared with the  blue histograms in the right panels of Figure \ref{fig:xus&h}.
This selection of quads returns $h=1.177^{+0.028}_{-0.171}$ for Model A, $h=0.995^{+0.142}_{-0.073}$ for Model B, $h=0.990^{+0.183}_{-0.035}$ for Model C, and
$h=0.991^{+0.016}_{-0.018}$ for Model D. 

For all models, the amount of scatter has decreased, most drastically for Model D and only marginally for Model B.
It follows that a quad which probes a range of radii has less
freedom in the fitting process and correspondingly less scatter in $h$. For Model A, the median has changed substantially, while for the other models
the median has changed at the 1-2\% level.
It is unclear why the quads which probe a range of radii in Model A would be more biased than the other models, but is likely related
to the fact that Model A has the profile with the most drastic curvature i.e. departure from the power-law model 
(visible in Fig. \ref{fig:FourModelProfile}). The utility of making this selection in real surveys is questionable 
unless this biasing effect can be understood.

\subsection{Limitations to this study}\label{ssec:limits}
There are clear limitations to this study. This has been a preliminary investigation using simple analytical profiles as a stand-in for real galaxies. While exploring these simple cases in a controlled setting is valuable, only four variants on a similar profile have been tested, hardly enough to draw sweeping conclusions about all mass distributions. 

Comparing this work to \citet{Tagore18}, more quads were successfully fit with small $\chi^2$, but our work uses simple elliptical profiles with no lens environments or other complications. Discrepancies from an elliptical shape are prevalent in real lenses \citep{WW15,Gomer18}, although the effect of such complexities on the recovery of $h$ is unknown. This topic will be further explored in a coming paper (\citet{Gomer20}, submitted for review). \citet{Tagore18} also examined mock lenses over different redshifts, while all lenses in this study are at the same redshift. 


The probability distributions used in the MLE for this work also have a subtle limitation, as they are constructed from point estimates of individual fits, marginalized over all parameters except $h$. This is not strictly equivalent to the posterior probability coming from an MCMC sampling of the complex parameter space, as would be done in a Bayesian analysis like that of H0LiCOW. If there are differences in the locations of these distributions at the 1\% level, this distinction could be relevant. More thorough studies in the future should be sure to invest computational resources into MCMC sampling of the distributions.

Finally, our interpretation regarding the slope constraint is that stellar kinematic constraints are equivalent to holding the slope at the weighted average value over the radii of the kinematic measurement. If our understanding is correct, the mismatch between the slope corresponding to an unbiased $h$ and the actual slope at the probed radius warrants skepticism about the process of using kinematic constraints to break the MSD. This finding is supported by our test using spherical Jeans arguments, but this interpretation needs to be confirmed. The next logical step is a study which incorporates kinematics into the fitting in a way which truly matches the H0LiCOW analysis, but is done for synthetic lenses where the deviation from the model profile is known. 

Perhaps the largest difference between this study and that of H0LiCOW is our use of point sources. Extended sources add additional information to the fitting process, although their constraints are not necessarily unique \citep{Saha01} and the level to which they can help with degeneracies is debated \citep{Walls18}. Nonetheless, the inclusion of extended images is necessary to have a more apt comparison to the H0LiCOW analysis. Until such a confirmation study can be done, caution is justifiable regarding our slope interpretation. This is especially relevant given that one of our main findings is that lensing degeneracies are less predictable than our intuition implies. 

\section{Conclusion}
Gravitational lensing is a competitive method for measurement of $h$ to 1\% precision independent of the distance ladder or the CMB. To reach this goal, degeneracies inherent to lens modeling must be precisely quantified and accounted for. To explore the effects of lensing degeneracies on $h$ recovery, we constructed quad lens systems from a series of two-component profiles, then fit these quads with a model different from the true profile: a power-law model. We then determined recovered distribution of $h$ values and compared them to the analytical predictions of \citet{Xu16}.

Our first finding is that the bias (location of the median) of the distribution of $h$ does not correspond to the value of $\lambda$ predicted by the mass-sheet transformation arguments in \citet{Xu16} and \citet{Tagore18}. Lensing degeneracies more complicated than the MSD \citep{Saha06} have conspired in an unexpected way to return unanticipated values of $h$. Since the result did not match the predicted value, we are skeptical about the existence of a straightforward calculation which could convert directly from a profile shape to the bias on $h$. Instead, the distribution of the bias on $h$ can only be reliably determined via the creation of mock quads fit with software. 

We also explore the effect of the inclusion of stellar kinematics by constraining the slope in the fitting process, which emulates the process by breaking the MSD through the inclusion of external information. We find that when the slope is held to the true value of the slope near the Einstein radius, $h$ can be considerably biased (23\% for Model A, 0-8\% for Models B,C, and D), depending on the exact bounds over which the slope is considered. Strangely, the addition of the correct information has caused the fitting to return an incorrect value. The value of slope which results in no bias on $h$ does not correspond to the true slope, perhaps indicating that ``slope'' acts as fitting parameter rather than describing the physical slope of the density profile. The inclusion of kinematics breaks the degeneracy, but can do so incorrectly, so as to introduce a significant bias. 

A remarkable consistency across all four of our models is that when the inner radius used in the determination of slope, $r_1\simeq0.15R_E$, the calculated slope results in zero bias in $h$, insensitive to the outer radius, $r_2$. If the spatial resolution of kinematic surveys can be increased to probe this region, the constraints placed by such measurements would not introduce a bias on $h$. At present such inner radii are out of reach. It may be possible to explore this region through simulations, although the resolution of modern simulations is insufficient, with $0.15R_E\simeq0.9\epsilon$ in the Illustris or EAGLE simulations.

A further consistency across our four models is that the slope which corresponds to an unbiased recovery of $h$ matches the density-weighted measure of slope within the Einstein radius. Furthermore, this value matches the local slope of the profile at $\simeq0.3R_E$ for all four profiles. If this slope could be measured, it could be used to inform the fitting process and recover an unbiased $h$, although it is not directly observable.

Apprehension regarding stellar kinematic constraints is supported by our examination of spherical Jeans kinematic information. Comparison of the projected integrated velocity dispersion between the actual profile and that of the power-law models found that the actual constraint did not correspond to models with the unbiased value of $h$. To force the fitting to match the stellar kinematics would pull the fit away from the correct value of $h$.

One interpretation of this result is that it may be preferable to not fix the slope or use kinematics if the only goal is a minimally biased value of $h$. Unbroken degeneracies will increase the scatter of the distribution, but may not bias the recovery as drastically as constraining slope to the incorrect value would. We suggest future studies carefully consider the potential pitfalls of biases inherent to the inclusion of stellar kinematics.

Our findings support the example of \citet{Kochanek19}, where a simplified lens model with stellar kinematic constraints can return a value of $h$ which is biased by more than the claimed H0LiCOW precision. We caution, however, that quantitative claims about $h$ based on profile shape may suffer the same shortcomings as the $\lambda$-based calculations of \citet{Xu16}.

Finally, we were motivated to search for an observable selection criterion which could reduce either the bias or scatter in $h$. We cannot confirm the findings of \citet{Tagore18} that cusp/fold/cross orientations have an effect on the recovery of $h$, but we do find noticeable reduction in scatter for quads with images which span a greater range of radii (Fig. \ref{fig:scattertest}). When limiting our sample to quads with $\Delta r/R_E\ge0.2$, the scatter is reduced in all cases. We note that this selection introduces substantial bias in the case of Model A, which is the model most different from a power law. This bias merits caution with respect to the utility of this selection in real surveys.

We would like to conclude by saying that lensing degeneracies are a subtle and treacherous reality. Their numerous manifestations are hidden behind high dimensional fitting processes, making them difficult to parse. The reasoning of \citet{Xu16} appears solid, and yet the prediction does not match reality. Our interpretations regarding stellar kinematic constraints may too be flawed in some deeper way. The way forward must be through the creation of mock systems complete with stellar kinematic models consistent with the methodology of observational studies. A major challenge is that these additional complications introduce even more parameters for degeneracies to lurk within. These degeneracies must be tackled in order to reliably constrain $H_0$ to the 1\% level.

\appendix
\section{Consistency Checks}\label{sec:conchecks}
As test for robustness, we run a few alterations to our fitting to confirm the resulting distributions of $h$ are unaffected by
our particular fitting process. These alterations are done with respect to Model D test with the slope allowed to vary, 
with the anticipation that they will apply similarly to all other fittings in this paper.
We describe these tests in detail here. We did not perform the MLE determination of $h$ for these tests, 
so results should be compared with the blue histogram in Figure \ref{fig:xus&h} for Model D.

The first alteration happens on the very first step of the fitting procedure described in Section \ref{sec:prelim}, where the ellipticity and
shear are held at 0.1 to search over the values for position angle and shear angle. This initialization value was set to 0.1 for 
both shear and ellipticity, which is defined in \texttt{lensmodel} as $1-q$. Since this value is an arbitrary choice on our part, we decided to test the effect of
increasing it to 0.3, a value more extreme than in any of the lenses. The resulting distribution of $h$ is not measurably different from the unaltered Model D,
with $h=0.969^{+0.040}_{-0.173}$, $f_{\chi^2/dof}<1=0.99$, and $R_{ell}=1.00$. When a KS test is performed to compare the $h$ distributions, 
the p-value is 73\%.

The second modification is to change the bounds over which the shear/ellipticity grid search is performed in the second run.
The unaltered version searches over values between 0.0 and 0.4 for both ellipticity and shear. This test instead searches over more extreme values which
are not consistent with zero, from 0.1 to 0.6. Again the goal is to show that even if one makes extreme choices in the fitting setup, the results are robust.
Again the resulting distribution of $h$ is the same as the unaltered Model D,
with $h=0.975^{+0.020}_{-0.218}$, $f_{\chi^2/dof}<1=0.94$, $R_{ell}=1.00$, and a p-value of 67\%. 

One final test of robustness is performed. This time we are curious not about the fitting initialization parameters, but about whether 100 quads is a sufficient number to 
accurately determine the distribution of $h$. We therefore run one test for Model D which is the same as the unaltered test except that it has 500 quads instead
of 100. The result is a distribution with $h=0.973^{+0.029}_{-0.168}$, $f_{\chi^2/dof}<1=0.95$, $R_{ell}=1.00$, p-value of 99\%.

The distributions across these tests are indistingushable. We therefore conclude that our distributions of $h$ are not significantly 
affected by either our fitting procedure or by small-number statistics.
The main quantity of interest, the median of $h$, varies by 0.006 across these tests, which is
less than the 1\% benchmark to which we desire accuracy.

\section{Single Quad Fitting}\label{sec:singlequad}
\begin{figure}
 \centering
   \includegraphics[width=\linewidth]{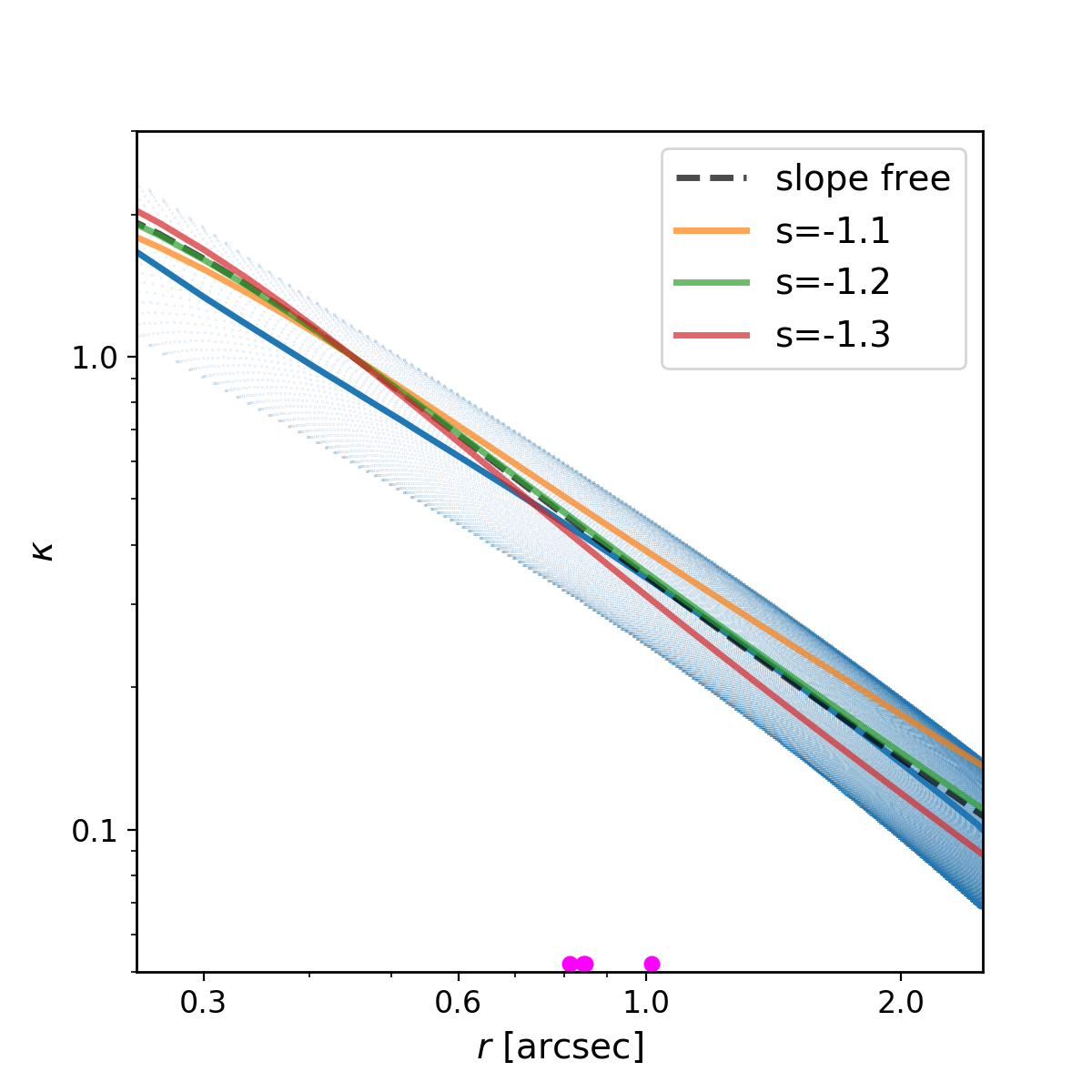}
   \caption{Convergence as a function of radius for the Model D input lens (blue) and several power-law fits to the image positions and time delays. Because 
       the lens is elliptical, the density takes on a range of values at any given radius. The circularly averaged profile is depicted by the solid blue line. 
       When the slope is free to vary,
       the dashed black curve is recovered, while the solid yellow, green, and red curves correspond to the slope being held at -1.1,-1.2, and -1.3, respectively. The magenta points
       near the horizontal axis depict the radial positions of the images. For comparison with Figure \ref{fig:FourModelProfile}, 
       1 arcsecond corresponds to approximately 6.7 kpc, varying slightly depending on the recovered value of $h$ for a given model.
        }
\label{fig:singquadkappa}
\end{figure}

Here we present a thorough analysis of a single quad from Model D, fit with different values of slope. This particular quad returns a nearly unbiased value of $h$ 
when the slope is free to vary (0.991 relative to 1.0). For comparison with Figure \ref{fig:scattertest}, $\theta_{23}=56^{\circ}$ and $\Delta r/R_{E}=0.24$.
Figure \ref{fig:singquadkappa} shows the mass density as a function of radius for the synthetic lens as well as several different 
fits to the image postitions and time delays. All fits lie within the scatter of the points, but the fit that matches the true profile best
is that when the slope is free to vary. 
In this case, the recovered slope is $-1.214$.
When the slope is held, the results are as follows, with the relative bias on $h$ in parentheses: $s=-1.1$ (0.902), 
$s=-1.2$ (0.986), $s=-1.3$ (1.005).

\begin{figure}
 \centering
   \includegraphics[width=0.6\linewidth]{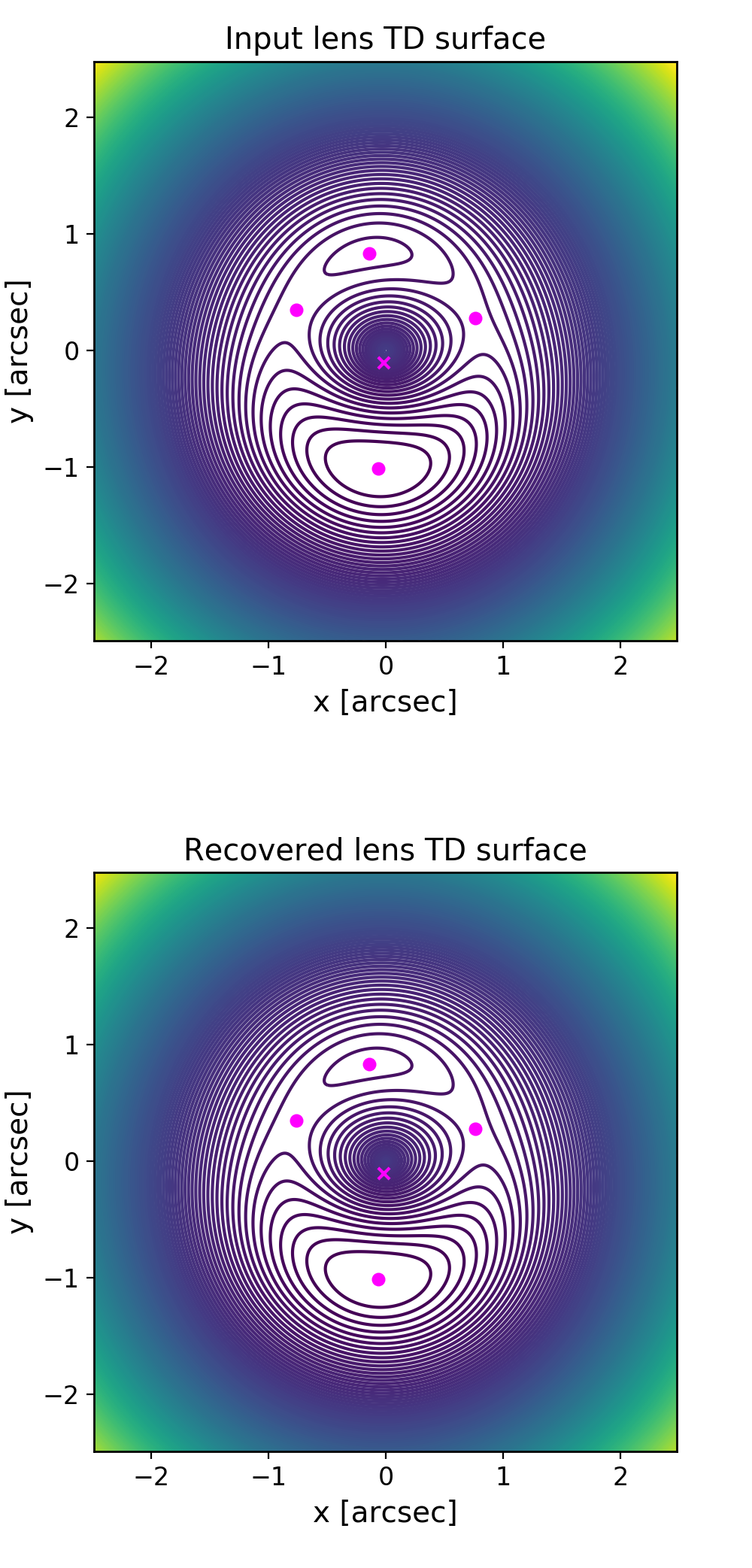}
   \caption{Time delay surfaces for the input lens/quad (top) and the recovered model lens/quad for the fit where the slope is allowed to vary(bottom).
           The image positions (magenta points) are recovered very well and the shape of the time delay surface matches very closely.
           }
\label{fig:singquadTDcompare}
\end{figure}

\begin{figure}
 \centering
  \begin{tabular}[c c]{cc}
   \begin{subfigure}[c]{0.45\linewidth}
    \includegraphics[width=\linewidth]{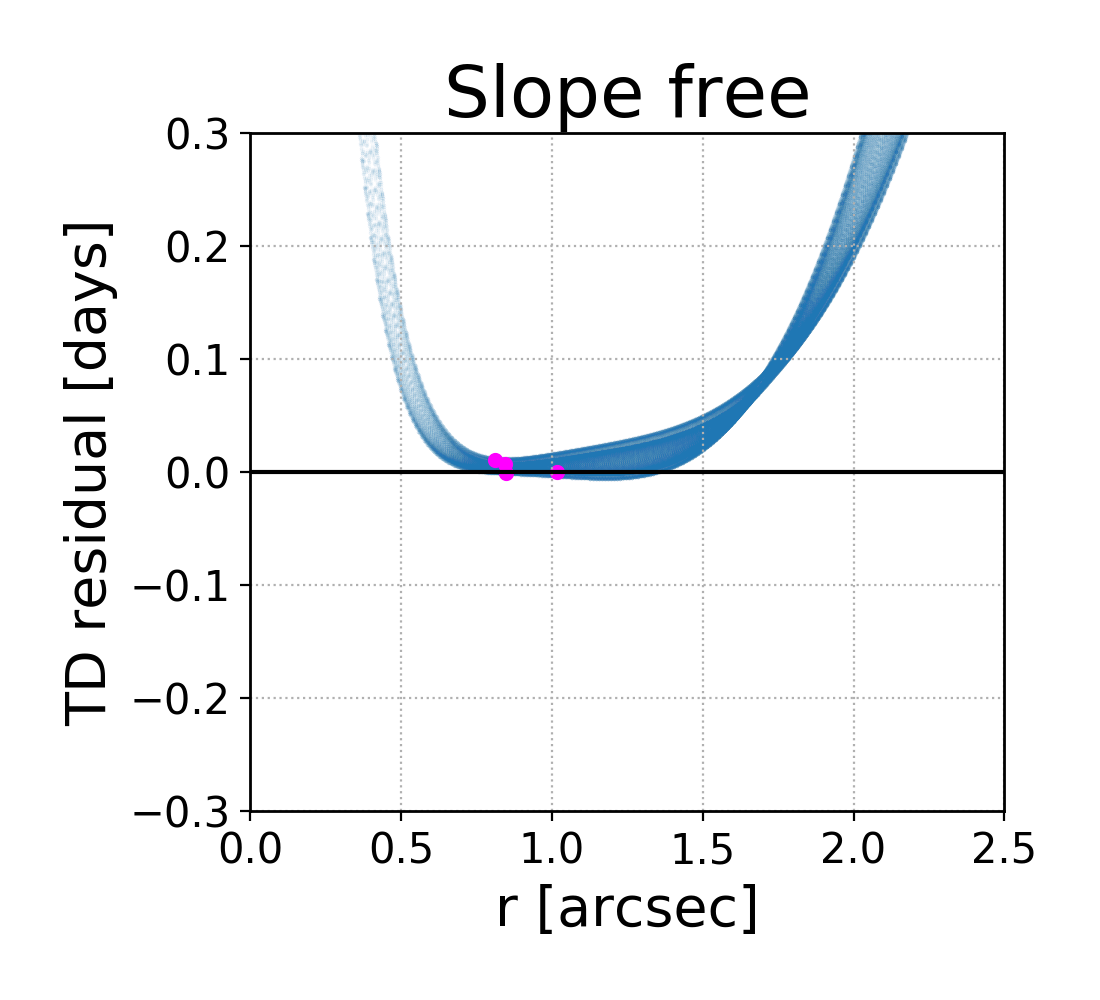}
   \end{subfigure} &
   \begin{subfigure}[c]{0.45\linewidth}
    \includegraphics[width=\linewidth]{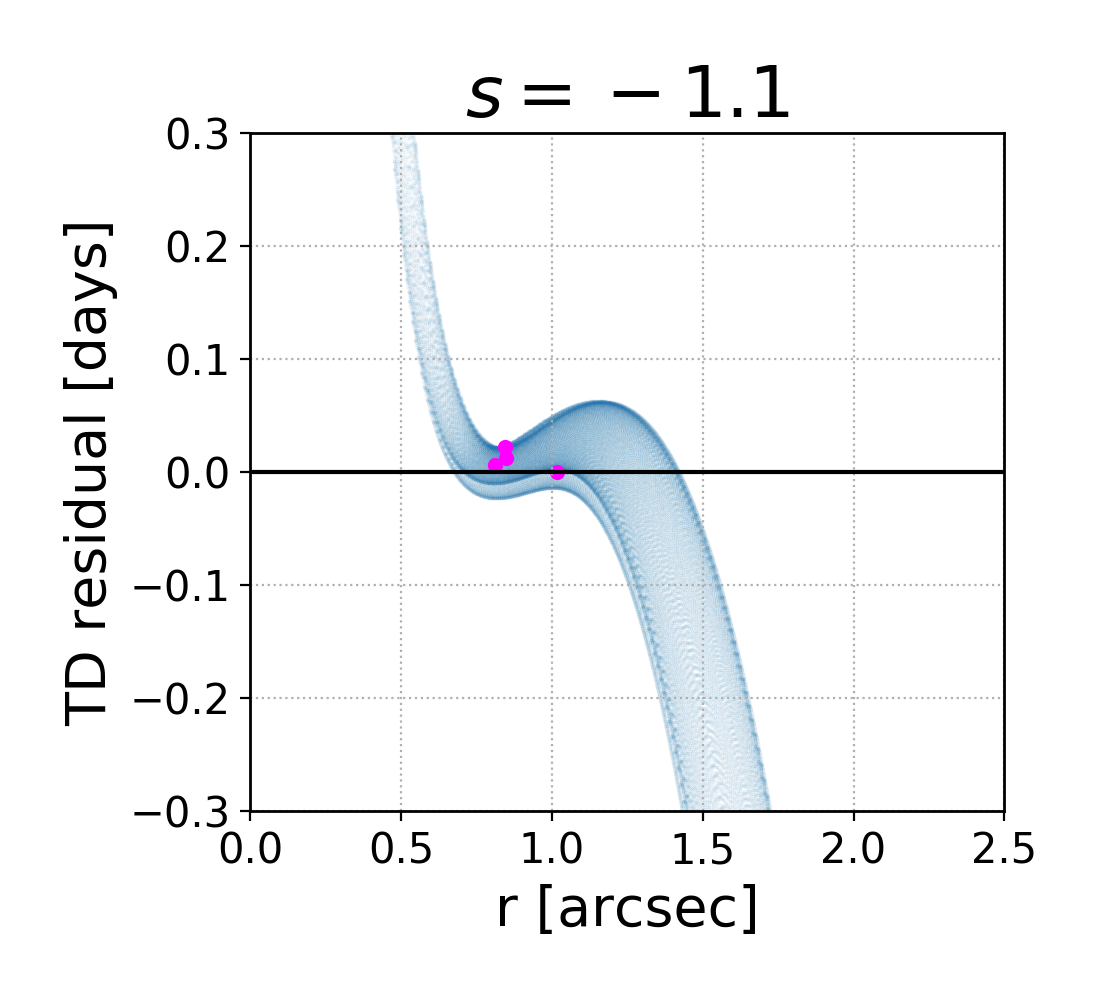}
   \end{subfigure} \\
   \begin{subfigure}[c]{0.45\linewidth}
    \includegraphics[width=\linewidth]{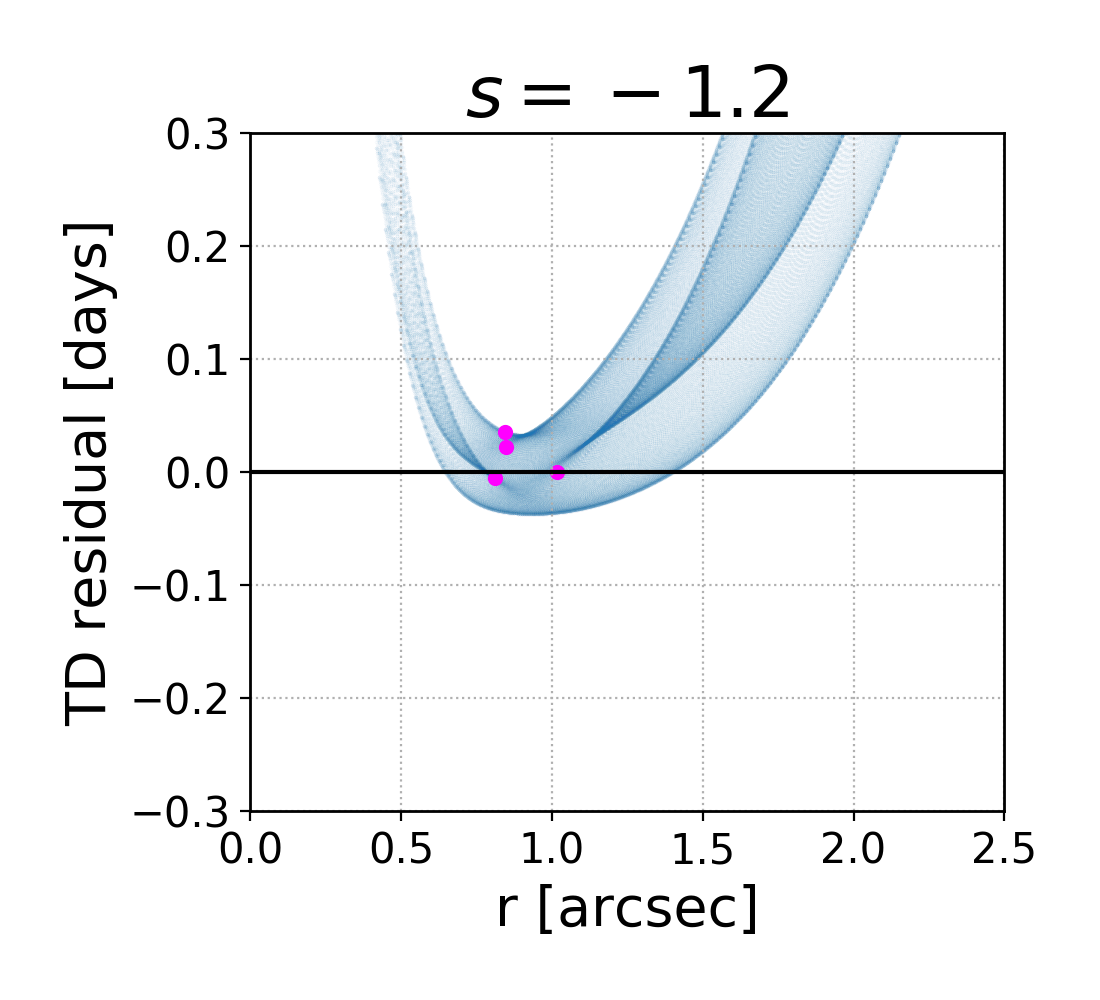}
   \end{subfigure} &
   \begin{subfigure}[c]{0.45\linewidth}
    \includegraphics[width=\linewidth]{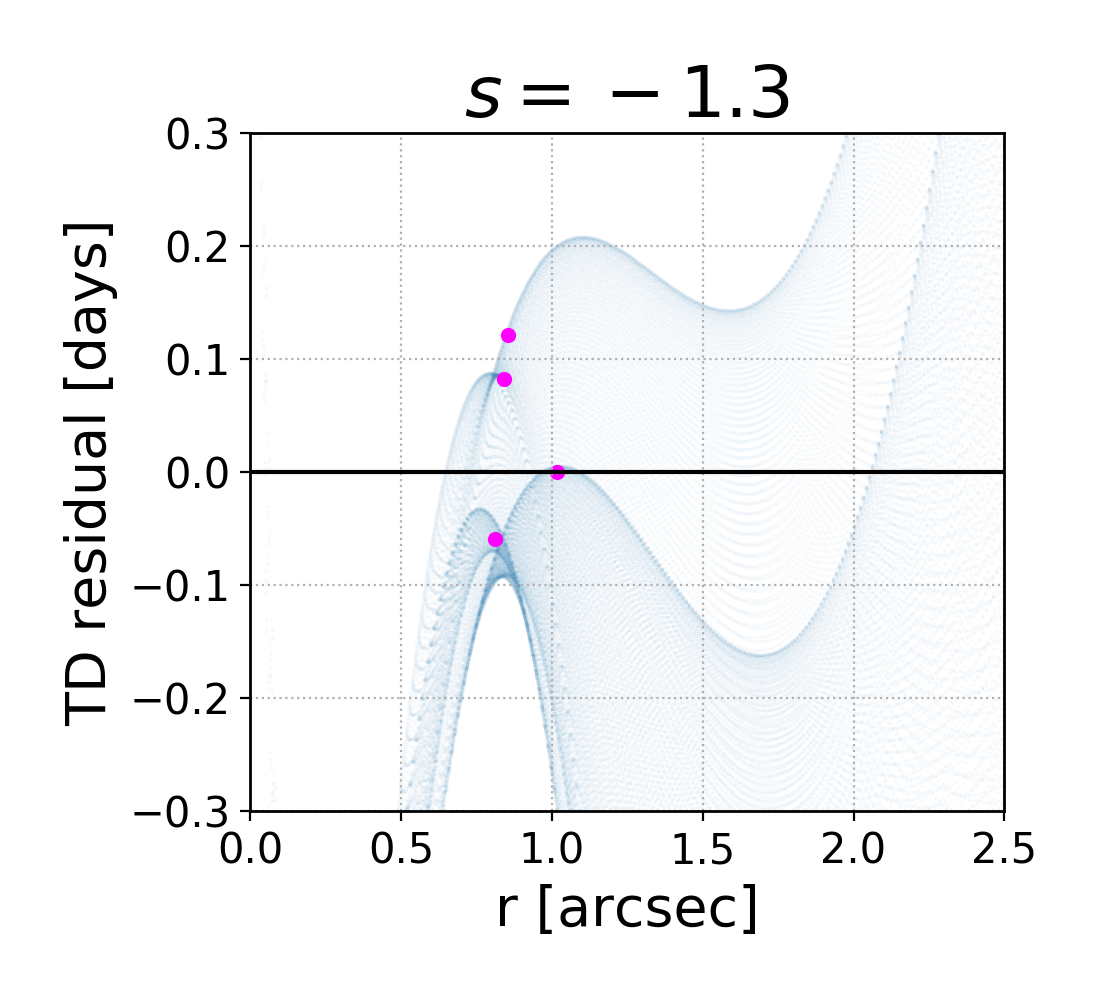}
   \end{subfigure} 
  \end{tabular}
\caption{Residuals of the fit time delay surface relative to the input lens. In the top left panel, each blue point corresponds to a single pixel of
         Figure \ref{fig:singquadTDcompare}, where the slope is free to vary in the fitting procedure. For each pixel, the difference has been evaluated between input surface and the fit surface 
         (top and bottom panels of Fig. \ref{fig:singquadTDcompare}) and plotted now as a 1D function of radius. The image locations themselves are represented
         as magenta points. The arbitrary offset of lensing potential is set such that the value for the first-arriving image matches across the two models.
         The fit surface matches the input quad very well, especially for radii between $0.5''$ and $1.75''$,
         evidenced by the fact that the residual in this region is nearly zero (the time-delay uncertainty in the evaluation of $\chi^2$ is 0.1 days).
         The remaining panels show the fits for the three cases where the slope is held at a particular value. 
         As evident by the larger vertical spread in the residuals, these fits do not match the actual
         surface as well as the case where the slope is free to vary. Nonetheless, the vertical spread of the images is fairly small, 
         confirming that the time delay differences between the images are close to their correct values 
         (within about 1.2$\sigma$ in the worst case, where slope is fixed at -1.3).
         }
\label{fig:singquadTDresidual}
\end{figure}

Figure \ref{fig:singquadTDcompare} compares the time delay surfaces of the true input quad (top) and the best-fit result from the case where the slope is free to vary.
The surface is reproduced well, with images and time delays matching the quad too accurately to dicern by eye
 ($\chi^2/dof\simeq10^{-3}$). To explore this, Figure \ref{fig:singquadTDresidual} plots the residual difference between the true lensing potential and the fit potential, now
as a 1D function of radius, for each of the four fits from Figure \ref{fig:singquadkappa}. Since potentials allow for an arbitrary choice of vertical offset, we choose
to set the comparison to equate the first-arriving image.
A good match would be represented by the points being laid out in a flat surface with nearly zero residual. Since $\chi^2$ is calculated with an uncertainty
of 0.1 days, as long as the surface residuals are within 0.1 days of zero, the time-delay $\chi^2$ will be small. 
The closest match of the four fits is the case where the slope is allowed to vary, which closely matches the time-delay surface between $r=0.5''$ and $r=1.75''$.
All fits result in the image time delays being less than $1\sigma$ from the true values, except the case where slope is held at -1.3, 
which has one image off by $\simeq1.2\sigma$.

The images and time delays of this quad are well-recovered by the fitting procedure, instilling confidence that the results for the large set of quads are reliable.

\section{Scaling relations of velocity dispersion constraints}\label{sec:analyticalsigma}
In Section \ref{ssec:jeans}, it was shown that the contours of stellar velocity dispersion run nearly parallel to the MSD-$b$-$\alpha$ relation. 
This somewhat counterintuitive result implies the stellar velocity dispersion can do little to help with the MSD-- the stellar kinematics cannot provide additional constraints to those which have
already been provided by the lensing information. 
This relationship is actually fairly straightforward to derive using simple scaling relations for an isotropic power law, without the use of any numerical fitting procedure. 
The question is: does the constraint from the integrated stellar velocity dispersion within some aperture radius provide a unique constraint from that of
enclosed mass at the Einstein radius?
To begin, let us assume both the density and $\sigma^2$ scale as power laws with radius, with slope $\gamma$
\footnote{Note that $\gamma$ is defined as positive. To compare with the 2D lensing potential profiles in the text, which use $\alpha$, $\gamma$=3-$\alpha$.} 
for density and some unknown slope $\beta$ for velocity dispersion:
\begin{equation}
\rho=\rho_0 \left( \frac{r}{r_s} \right) ^{-\gamma}
\end{equation}
\begin{equation}
\sigma^2=c \left( \frac{r}{r_s} \right) ^{-\beta}
\end{equation}
The mass enclosed within a radius $R$ is then
\begin{equation}
M(R)=\int_{0}^{R}4\pi \rho r^2  dr = \frac{4\pi \rho_0 r_s^3}{3-\gamma} \left( \frac{R}{r_s} \right) ^{3-\gamma}
\end{equation}
From Jeans hydrostatic equilibrium,
\begin{equation}
\frac{d}{dr}(\rho \sigma^2)= -\rho \frac{GM}{r^2}
\end{equation}
\begin{equation}
\frac{d (ln(\sigma^2))}{d (ln (r))} +\frac{d (ln(\rho))}{d (ln (r))} = -\frac{GM}{\sigma^2 r}
\end{equation}
\begin{equation}
\beta +\gamma = \frac{4\pi G \rho_0 r_s^2}{c(3-\gamma)} \left( \frac{r}{r_s} \right) ^{2-\gamma+\beta}
\end{equation}
Because the left hand side is constant for a given system, we can conclude $\beta=\gamma-2$. Our knowledge of the isothermal case supports this-- a density slope of 2 corresponds to a constant velocity dispersion.
We can also determine the velocity dispersion normalization, $c$, from this equation. Finally, we can calculate the weighted velocity dispersion within an aperture radius $R_{ap}$ as:
\begin{equation}
\langle \sigma^2 \rangle = \frac{\int_{0}^{R_{ap}} 4\pi \rho \sigma^2 r^2 dr}{\int_{0}^{R_{ap}}4\pi \rho r^2  dr}=\frac{2\pi G \rho_0 r_s^2 }{(\gamma-1)(5-2\gamma)} \left( \frac{R_{ap}}{r_s} \right) ^{2-\gamma}
\end{equation}

\begin{figure*}
 \centering
   \includegraphics[width=\linewidth]{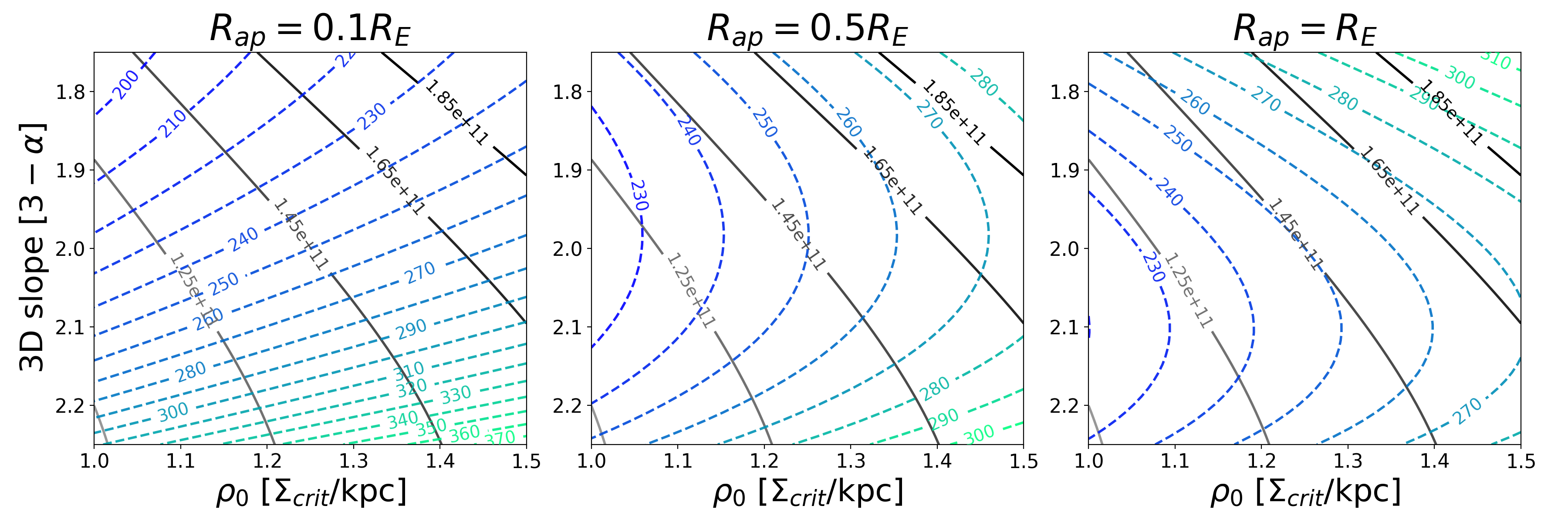}
   \caption{For the isotropic power-law density profile, the enclosed mass (in solar masses, grayscale contours) and 
    integrated weighted stellar velocity dispersion (in km/s, blue-green dashed contours) can be 
    readily calculated given the density slope (y-axis) and normalization (x-axis). 
    For the three panels, different radii are used for the integration of the velocity dispersion, ranging from $0.1R_E$ to $R_E$.
    Because of the mass-sheet degeneracy, lensing alone is able to make a determination of only the enclosed mass within $R_E$, 
    giving no information about the slope (which can be thought of as a proxy for $h$ in this example).
    To break the degeneracy, stellar kinematic data is commonly used in the literature, but this can only provide useful constraints on the slope if the contours for velocity dispersion run at 
    approximately perpendicular angles to the constant-mass contours. In this example, the mass enclosed within the Einstein radius is $1.45\times10^{11}M_{\odot}$. Consider
    an integrated stellar velocity dispersion constraint of 250 km/s (corresponding to the isothermal case) and observe the intersection of these two contours. 
    For an aperture radius of $0.1R_E$, a measurement of integrated velocity dispersion could provide useful constraints, 
    but for an aperture radius of similar order to $R_E$, the contours run nearly parallel and are degenerate with each other. 
    Stellar velocity constraints cannot help break the MSD in this case, even under
    ideal circumstances where the model is exactly correct. Compare to the numerical lens profiles in Figure \ref{fig:sigmaph}, but 
    note that the range of both axes is extended here to show a wider range of behavior.
    }
\label{fig:analyticalgrid}
\end{figure*}

We now see that the lensing constraint on mass within $R_E$ will scale as $R_E ^{3-\gamma}$ while the velocity dispersion measurement will scale as $R_{ap}^{2-\gamma}$, with some $\gamma$ dependence in the normalizations.
How contours of constant mass compare with those of constant measured velocity dispersion will be dependent on the ratio of $R_{ap}/R_{E}$. In Figure \ref{fig:analyticalgrid},
we show three plots with different values of $R_{ap}/R_{E}$, and find that when the two are equal then the contours are nearly parallel, demonstrating
that the usefulness of the stellar kinematic constraint depends on the aperture size over which it is measured. This supports the numerical findings of this paper-- that kinematic constraints
are only useful if velocity dispersion is measured within a sufficiently smaller radius than the Einstein radius.

\bibliographystyle{unsrtnat}
\bibliography{slopepaperresubmit}
\end{document}